\documentstyle[11pt,aaspp4]{article}

\def\simlt{\lower.5ex\hbox{$\; \buildrel < \over \sim \;$}}
\def\simgt{\lower.5ex\hbox{$\; \buildrel > \over \sim \;$}}
\def\gsim{\;\rlap{\lower 2.5pt
\hbox{$\sim$}}\raise 1.5pt\hbox{$>$}\;}
\def\lsim{\;\rlap{\lower 2.5pt
   \hbox{$\sim$}}\raise 1.5pt\hbox{$<$}\;}

\def\spose#1{\hbox to 0pt{#1\hss}}
\def\lta{\mathrel{\spose{\lower 3pt\hbox{$\mathchar''218$}}
     \raise 2.0pt\hbox{$\mathchar''13C$}}}
\def\gta{\mathrel{\spose{\lower 3pt\hbox{$\mathchar''218$}}
     \raise 2.0pt\hbox{$\mathchar''13E$}}}

\def\a3b{(\alpha+3\beta)}

\def\be{\begin{equation}}
\def\ee{\end{equation}}

\newcommand{\beq}{\begin{equation}}
\newcommand{\eeq}{\end{equation}}


\lefthead{MENOU \& TABACHNIK}
\righthead{DYNAMICAL HABITABILITY}

\received{2002 August 30}
\begin{document}

\title{Dynamical Habitability of Known Extrasolar Planetary Systems}

\author{Kristen Menou\altaffilmark{1,2} and Serge Tabachnik}

\affil{Princeton University, Department of Astrophysical
Sciences, Princeton, NJ 08544, USA}

\altaffiltext{1}{Chandra Fellow}
\altaffiltext{2}{Present address: Department of Astronomy, P.O. Box 3818, 
University of Virginia, Charlottesville, VA 22903, USA}

\vspace{\baselineskip}

\begin{abstract}
Habitability is usually defined as the requirement for a terrestrial
planet's atmosphere to sustain liquid water. This definition can be
complemented by the dynamical requirement that other planets in the
system do not gravitationally perturb terrestrial planets outside of
their habitable zone, the orbital region allowing the existence of
liquid water. We quantify the dynamical habitability of 85 known
extrasolar planetary systems via simulations of their orbital dynamics
in the presence of potentially habitable terrestrial planets.  When
requiring that habitable planets remain strictly within their
habitable zone at all time, the perturbing influence of giant planets
extends beyond the traditional Hill sphere for close encounters:
terrestrial planet excursions outside of the habitable zone are also
caused by secular eccentricity variations and, in some cases, strong
mean-motion resonances.  Our results indicate that more than half the
known extrasolar planetary systems (mostly those with distant,
eccentric giant planets) are unlikely to harbor habitable terrestrial
planets.  About $1/4$ of the systems (mostly those with close-in giant
planets), including $1/3$ of the potential targets for the {\it
Terrestrial Planet Finder}, appear as dynamically habitable as our own
Solar System.  {The influence of yet undetected giant planets in
these systems could compromise their dynamical habitability. Some
habitable} terrestrial planets in our simulations have substantial
eccentricities ($e > 0.1$) which may lead to large seasonal climate
variations and thus affect their habitability.
\end{abstract}



\section{Introduction}

As of August 2002, high-precision radial velocity surveys have
revealed the presence of 97 low-mass companions around $85$ nearby
Sun-like stars (Marcy, Cochran \& Mayor 2000; see, e.g., the
extrasolar planet almanac\footnote{{\tt
http://exoplanets.org/almanacframe.html}} and
encyclopedia\footnote{{\tt
http://www.obspm.fr/encycl/encycl.html}}). The Jovian nature of these
companions was confirmed by the observation of transits in the system
HD~209458, which indicated the presence of a planet of mass $M_p =0.69
M_{\rm Jup}$ and radius $R_p = 1.35 R_{\rm Jup}$ (in Jupiter units;
Charbonneau et al. 2000; Henry et al. 2000; Mazeh et al. 2000). A
striking feature of the planetary systems discovered by these surveys
is their orbital configuration. In most cases, gaseous giant planets
are either located in the close vicinity of their parent star, on
nearly circular orbits, or further away from their star but then often
on orbits with substantial eccentricities (see Table~\ref{tab:one}
and~\ref{tab:onebis}). These orbital configurations are clearly
different from those in our own Solar System, where gaseous giants are
located at least several AUs from the Sun, on low eccentricity orbits.
The detection of giant planets with orbital characteristics comparable
to those of Jupiter has been only recently announced (Marcy et
al. 2002; Mayor et al. 2002).

It is reasonable to expect that extrasolar planetary systems
discovered to date were not only able to form Jovian planets but
probably also formed less massive, terrestrial planets (see, e.g.,
Ruden 1999). This raises the exciting possibility of the presence of
life on a ``suitable'' terrestrial planet in known extrasolar
planetary systems.  The detection and the determination of the
suitability for life of extrasolar planets around stars within $\sim
20$~pc of our Sun (including those already known to harbor Jovian
planets) is the scientific motivation behind efforts to build a
Terrestrial Planet Finder (TPF) within the next decade.\footnote{{\tt
http://planetquest.jpl.nasa.gov/TPF/tpf\_index.html}}

The suitability of a terrestrial planet to harbor life is usually
referred to as habitability, and the region around a star in which a
terrestrial planet's atmosphere is able to sustain the necessary
liquid water is labeled the habitable zone (see, e.g., Hart 1978;
Kasting, Toon \& Pollack 1988; Kasting, Whitmire \& Reynolds 1993 and
references therein; see \S2.2 for details). The concept of atmospheric
habitability can be complemented by a study of the orbital dynamics of
a planetary system in order to better evaluate its ``dynamical
habitability.''  Indeed, as opposed to Jupiter in our Solar System,
the gravitational zone of influence of many extrasolar giant planets
often approaches, or sometimes even overlaps with, the habitable zone
of their parent system. As a consequence, terrestrial planets
potentially located within these habitable zones could be
significantly perturbed and possibly driven out of the habitable zone
or the system altogether. It is the purpose of this study to quantify
the importance of these dynamical effects on the habitability of known
extrasolar planetary systems. Although dynamical effects on
potentially habitable planets have already been investigated by
several groups in the past (Gehman, Adams \& Laughlin 1996; Jones,
Sleep \& Chambers 2001; Laughlin, Chambers \& Fischer 2002; see also
Rivera \& Lissauer 2000, 2001), our study is different in the
formulation of the problem adopted and in that it addresses the entire
set of known extrasolar planetary systems. We compare our results with
those of previous studies in \S5.

In \S2, we describe the sample of extrasolar planetary systems used in
this work, the extent of their habitable zones and the size of the
zone of influence of each individual gaseous giant planet. In \S3, we
describe and justify the method of orbital integration adopted, while
the results of these integrations for all extrasolar planetary systems
are presented in \S4. Various consequences and implications of our
results are discussed in \S5.

\section{Definitions}

\subsection{Known Extrasolar Planetary Systems}

The sample of extrasolar giant planets used in this work is listed in
Table~\ref{tab:one} and~\ref{tab:onebis} (for the most recently
discovered planets), with relevant data on the distance to the system,
the stellar mass, the planet's mass (modulo the unknown orbital
inclination), semi-major axis, orbital period and eccentricity.  These
data were taken from the compilation of all published survey results
found in the extrasolar planet encyclopedia (see footnote 3), with
some modifications when required. Consistency of values listed in the
main catalog table with those given in tables specific to each system
was verified. Eventual discrepancies were corrected by going back to
the published values. We note that even orbital parameters estimated
for a same system by different groups sometimes show discrepancies, so
that values listed in Table~\ref{tab:one} and~\ref{tab:onebis} cannot
all be considered as being accurate. We have verified that
Table~\ref{tab:one} and~\ref{tab:onebis} are in very good overall
agreement with the corresponding table published by the California and
Carnegie Planet Search (in most cases, discrepancies are $\lsim 0.1$
for the eccentricity and $\lsim 10-20\%$ for other orbital
parameters). These uncertainties should not significantly affect our
conclusions, since we are mainly interested in the global statistics
of the entire set of systems.

It is worth mentioning here the limitations of current radial velocity
surveys (see, e.g., Table~1 of Tabachnik \& Tremaine 2002; { Cumming
et al. 2002). While the best Doppler shift precisions achieved are
$\sim 2-3$~m~s$^{-1}$, which would guarantee that a stellar reflex
motion of amplitude comparable to that of our Sun under the influence
of Jupiter ($\sim 13$~m~s$^{-1}$) can be detected, a second important
limiting factor for} radial velocity surveys is their time
coverage. The oldest surveys started about $15$~years ago, and since
in practice an orbital solution is considered robust only once the
planet has completed $\gsim 2$ full orbital revolutions, current
surveys are limited to the detection of planets located within
$4$-$5$~AUs of their parent star (see Table~\ref{tab:one}
and~\ref{tab:onebis}). Consequently, current surveys (with their
limited time-coverage) would only marginally reveal the presence of
Jupiter in an extrasolar planetary system identical to our Solar
System located several tens of parsecs away, as evidenced by the only
very recent detection of 55~Cnc~d and Gl~777A~b (Marcy et al. 2002;
Mayor et al. 2002). Saturn (and all the other planets in the Solar
System) would not have been detected by current surveys. This {is an
important} point to keep in mind when we later compare dynamical
habitability in our own Solar System to that in extrasolar planetary
systems (see \S\ref{sec:equivss}).

\subsection{Atmospheric Habitability and Habitable Zones}

\subsubsection{Habitable Terrestrial Planets}

Various definitions of habitability have been used in the past (see
Kasting et al. 1993 for a historical perspective), but it is now
generally agreed that the development of carbon-based life requires
liquid water in the atmosphere of a terrestrial planet over geological
timescales (Kasting et al. 1993 and references therein; Rampino \&
Caldeira 1994). This relatively well-defined atmospheric requirement
is not easily translated into orbital requirements for a planet around
a star of given luminosity, however, because the planetary temperature
is regulated by subtle atmospheric feedback mechanisms which are not
all fully understood. For example, Hart (1978) constructed a model of
planetary habitability which accounted for the evolution of the solar
luminosity and included the runaway greenhouse effect and the
planetary glaciation effect. Because these two effects are positive
feedback mechanisms,\footnote{The greenhouse effect, which raises
planetary surface temperatures, is itself stronger at higher
temperatures because water vapor becomes more abundant in the
atmosphere. This is the origin of a potential runaway. Similarly,
glaciation results in a larger coverage of the planetary surface by
snow/ice, which tends to reduce even more the planetary surface
temperature by reflecting a larger fraction of the insolation flux.}
this author found a rather narrow radial extent for the habitable
zone. By contrast, Kasting et al. (1993) showed that a much larger
size for the habitable zone could be obtained if one accounts for the
negative feedback mechanism of the carbon-silicate cycle (in
particular the possibility of releasing CO$_2$, a greenhouse effect
gas, through volcanic activity even after planetary glaciation
occurs).

The example above illustrates the subtleties involved in calculating
the radial extent of habitable zones. Even the calculations of Kasting
et al. (1993), that we use in the present work, are subject to
uncertainties. The origin of these uncertainties is in the approximate
treatment of H$_2$O clouds, the neglect of CO$_2$ clouds (which could
significantly increase the outer radius of the habitable zone through
a scattering variant of the greenhouse effect; Forget \& Pierrehumbert
1997) and, to a lesser extent, dependences on atmospheric abundances
(a CO$_2$/H$_2$O/N$_2$ atmosphere is assumed in their calculations,
but relative abundances different from those in Earth's atmosphere can
affect the extent of the habitable zone), on planetary mass (see
below) and on biological influences (see Kasting et al. 1993 for a
detailed discussion). These authors present their results on habitable
zones under three different sets of assumptions. We use their
intermediate case, corresponding to the ``atmospheric runaway'' and
``maximum greenhouse effect'' limits, by fitting the curves of radial
extent for the habitable zone as a function of stellar mass given in
their Fig.~15. The corresponding values for the inner and outer radii
of habitable zones, accurate to within $10\%$ given our simple fitting
procedure, are reported in Table~\ref{tab:one} and~\ref{tab:onebis}
for each system.

We note that most of the extrasolar planetary systems listed in
Table~\ref{tab:one} and~\ref{tab:onebis} have stars with low enough
masses ($\lsim 1.2 M_\odot$) that the difference between zero-age
main-sequence habitable zones and continuously habitable zones (which
account for the increasing stellar luminosity with age) are very small
for timescales $\lsim 1$~Gyr. For simplicity, we only use zero-age
main-sequence habitable zones here (which do not require us to specify
stellar ages for each system). This choice is consistent with the
limited integration time of $10^6$~yrs covered by our dynamical study,
which thus describes, in each system, the first million years after
giant planets stopped migrating and terrestrial planets were fully
formed. A study over longer timescales, with continuously habitable
zones, is beyond the scope of the present work.

Kasting et al. (1993) have also considered in their study the
possibility that terrestrial planets may have masses different from
that of Earth (the relevant parameter in their model being surface
gravity). The authors argued that a terrestrial planet should be
several times more massive than Mars (say $\gsim 0.3 M_\oplus$) to
retain its atmosphere over long geological timescales and to sustain
tectonic activity as required for the carbon-silicate cycle to operate
(see also Williams, Kasting \& Wade 1997). They showed that planets of
masses $\sim 0.1 M_\oplus$ and $10 M_\oplus$ have habitable zones
comparable in size to that of an earth-mass planet (to within $\sim 5
\%$). The upper limit of $10 M_\oplus$ roughly corresponds to the
onset of runaway gas accretion and gaseous envelope formation in
standard planetary formation scenarios (see, e.g., Ruden 1999). In
what follows, we will therefore use the habitable zone limits listed
in Table~\ref{tab:one} and~\ref{tab:onebis} even when considering
terrestrial planets with masses anywhere in the range $0.3 M_\oplus$
to $10 M_\oplus$.

We note that Gliese 876 (spectral type M4V, $M_* \simeq 0.32 M_\odot$)
is unique among the systems listed in Table~\ref{tab:one}
and~\ref{tab:onebis} in that a terrestrial planet located within the
habitable zone of this system would also likely be tidally-locked to
the central star (see Fig.~16 in Kasting et al. 1993). Joshi, Haberle
\& Reynolds (1997) have argued that the permanent day and (much
colder) night sides on the planet, resulting from its synchronous
rotation, may still allow the planet to be habitable because
atmospheric circulation can efficiently redistribute heat from the
day-side to the night-side and thus smooth out the large temperature
gradients expected from pure radiative equilibrium balance. Tidal
forces would also lead to circularization of the terrestrial planet
orbit, an effect that may help retain terrestrial planets within the
habitable zone but is not included in our dynamical study of this
system.

\subsubsection{Habitable Moons of Giant Planets}

Williams et al. (1997) have proposed that moons around extrasolar
giant planets located within a system's habitable zone could
themselves be habitable. As for terrestrial planets, the moon must be
massive enough ($\gsim 0.2 M_\oplus$ in general) to both retain its
atmosphere over long timescales and sustain the tectonic activity
essential to the carbon-silicate cycle. Another important requirement
for this moon is that it must possess a strong enough magnetic field
to protect its atmosphere against the bombardment of energetic ions
from the giant planet's magnetosphere. Altogether, these are rather
stringent requirements which nonetheless permit the existence of
habitable moons in extrasolar planetary systems.

Williams et al. (1997) also noted the additional difficulty existing
for habitable moons when large variations of the insolation flux are
caused by the Jovian planet's substantial eccentricity. We identify 19
cases in Table~\ref{tab:one} and~\ref{tab:onebis} for which the giant
planet's semi-major axis is located within the system's habitable
zone. However, only 5 such planets have eccentricities small enough to
remain within the habitable zone during their entire orbital
revolution (indicated by stars in the last column of
Table~\ref{tab:one} and~\ref{tab:onebis}). These are the planets which
are most likely to harbor habitable moons. We note that these five
planets are located far enough from their parent star (periastron
distance $>0.6$~AU) that the mass of their potential moons are not
dynamically constrained (Barnes \& O'Brien 2002). {Another possibility
for the habitability of (distant) moons, that we will not discuss
further here, is when the right conditions for the development of life
are provided by tidal heating (as is thought to be the case on
Europa).}

\subsection{Zone of Influence and Classes of Dynamical Habitability}
\label{sec:class}

The definition of a planet's gravitational zone of influence often
used in dynamical studies is 3 times its Hill radius,
\begin{equation}
R_{\rm Hill} \equiv a \left( \frac{M_p}{3M_{*}} \right)^{\frac{1}{3}},
\end{equation}
where $a$ is the planet's semi-major axis, $M_p$ is the planet's mass
and $M_{*}$ is the mass of the central star. Three Hill radii cover
the region of close encounters with the planet, which typically result
in collisions with the planet, the central star, or ejections from the
system. Since a planet with finite eccentricity, $e$, experiences
radial excursions from $(1-e)a$ to $(1+e)a$, we generalize the
definition of the planet's zone of influence to include the entire
region extending from $R_{\rm in}=(1-e)a-3R_{\rm Hill}$ to $R_{\rm
out}=(1+e)a+3R_{\rm Hill}$. The values of the inner and outer radii of
the zone of influence of all extrasolar giant planets are reported in
Table~\ref{tab:one} and~\ref{tab:onebis} (a planetary mass
corresponding to $\sin i =0.5$ was specifically assumed for these
numerical estimates).

Based on the degree of overlap between the planet's zone of
influence (ZI) and the system's habitable zone (HZ), we define 4
classes of dynamical habitability:
\begin{itemize}
\item Class I: $a \leq 0.25$~AU 
\item Class II: $a > 0.25$~AU and no overlap between HZ and ZI 
\item Class III: $a > 0.25$~AU and partial overlap between HZ and ZI
\item Class IV: $a > 0.25$~AU and HZ is fully inside ZI
\end{itemize}
Class I is defined to include close-in extrasolar giant planets which,
because of their proximity to the parent star, should not
gravitationally influence much planets in the habitable zone.  The
classes of all extrasolar giant planets are reported in the last
column of Table~\ref{tab:one} and~\ref{tab:onebis}. When two classes
are given for a giant planet, they indicate the range of possible
classes for the five values of inclination considered in our study
(see \S3.4). In systems with multiple giant planets, dynamical
habitability will be determined by the planet which most strongly
influences the habitable zone (i.e. of highest class, according to the
definitions above). Accordingly, we also list in the last column of
Table~\ref{tab:one} and~\ref{tab:onebis}, in parenthesis, the highest
dynamical class found among all planets in multiple planet systems.

Because of the complete overlap between zone of influence and
habitable zone, systems containing class-IV giant planets are unlikely
to harbor habitable terrestrial planets.\footnote{{We note that it is
in principle still possible for these systems to harbor habitable
terrestrial planets at the stable Lagrange points of their giant
planets. This possibility may be limited to the five giant planets
identified in \S2.2.2 as good potential hosts for habitable moons.}}
It is more difficult to estimate the likelihood of finding habitable
terrestrial planets in systems containing class-III giant planets,
however, because the overlap is then only partial. In the next
section, we describe how the dynamical habitability of all the systems
listed in Table~\ref{tab:one} and~\ref{tab:onebis} can be quantified
with detailed numerical simulations.

\section{Orbital Integration Method}
\label{sec:im}
\noindent

We perform a large number of integrations of the orbital dynamics of
extrasolar planetary systems that include the presence of potential
terrestrial planets within their habitable zones. For each system, the
habitable zone is initially seeded with 100 terrestrial planets
randomly distributed in semi-major axis. As a measure of dynamical
habitability, we use the statistics of the fraction of these planets,
treated as test particles, that remain in the habitable zone after a
fixed amount of integration time. In \S\ref{sub:nis}, we describe our
integration scheme. We validate the test particle approximation for
terrestrial planets in \S\ref{sub:tp} and we justify our choice of
initial conditions for the integrations in \S\ref{sub:ic}.

\subsection{Numerical Integration Scheme}
\label{sub:nis}

We can safely rely on the second order mixed variable symplectic (MVS)
integrator described by Wisdom and Holman (1991) for planetary systems
with a single companion. For multiple-planet systems, Saha \& Tremaine
(1992; 1994) refined the integration scheme by adding individual
planetary time steps which reduce considerably the CPU integration
time (for the Solar System, the CPU gain is $\sim 50\%$). Both methods
are tailor-made for long numerical integrations of planetary systems
in which the central body is the dominant gravitational influence
since symplectic properties prevent any spurious drift in
Poincar\'{e}'s invariants caused by truncation errors. All our
simulations include leading-order post-Newtonian relativistic
corrections (Saha \& Tremaine 1994), which are not negligible,
especially for systems with close-in giant planets.

The integration time step is chosen to be $1/12$ of the shortest
orbital period of all the planets in the system, which is either that
of the closest-in giant planet or that of an hypothetical test
particule with semi-major axis right at the inner edge of the
habitable zone. This limiting value was empirically selected to ensure
that the relative energy error has a peak amplitude of $\simeq
10^{-6}$ over the duration of the integrations. In systems with
multiple giant planets, the ratio of respective time steps for giant
planets is always taken to be an integer. For instance, the time step
of Ups And b is 4.617/12=0.38 days while the time steps for the other
two planets are in the ratio 1:50:200 (closest integer ratio
satisfying the $1/12$ requirement for the individual timestep of each
additional planet).

As integration proceeds, the orbits of terrestrial planets are
carefully monitored. A terrestrial planet is removed from the
integration if it meets one of the three following criteria:
\begin{itemize}
\item the terrestrial planet's orbit became parabolic or hyperbolic
(as determined by an osculating eccentricity $\geq 1$)
\item the terrestrial planet experienced a close encounter with a
giant planet (if it approached within $3R_{\rm Hill}$ of the giant)
\item the terrestrial planet crosses the limits of the system's
habitable zone
\end{itemize}
The first condition corresponds to the terrestrial planet being
ejected from the system altogether or colliding with the central
star. The second condition is chosen because a close encounter almost
certainly results in ejection from the system or collision with the
giant planet or the central star.

The third condition is chosen so that a terrestrial planet is
considered habitable only if it remains strictly within the habitable
zone during the entire integration. Our main motivation to adopt this
conservative definition of dynamical habitability (as opposed to the
requirement that the planet's semi-major axis remains within the
habitable zone limits; see, e.g., Jones et al. 2001) is that it is
presently unclear whether an excursion outside of the habitable zone,
even if temporary, will allow the planet to remain habitable. In
particular, we note that the radiative time constant of the earth's
atmosphere (over continental regions) is $\tau_{\rm rad} \sim c_p P_0
/ g 4 \sigma T_{\rm rad}^3 \sim 1$~month, where $c_p$ is the heat
capacity at constant pressure, $P_0$ is the atmospheric surface
pressure, $g$ is the gravitational acceleration, $\sigma$ is the
Stefan-Boltzmann constant and $T_{\rm rad}$ is the radiative
equilibrium temperature (the radiative time constant for Mars is even
shorter, $\sim $ a few days). For $\tau_{\rm rad} \ll P_{\rm orb}$,
one expects the atmosphere to respond efficiently to seasonal
variations of the absorbed stellar flux, which could potentially
damage the planet's habitability during excursions outside of the
habitable zone.\footnote{The problem is further complicated by the
fact that the radiative time constant of the atmosphere is larger by
1--2 orders of magnitude over large-heat-capacity oceanic regions
(see, e.g., Williams \& Kasting 1997; North, Mengel \& Short
1983).}. We postpone a more detailed discussion of the orbital
requirements for dynamical habitability to a future publication and we
adopt here a conservative definition of dynamical habitability which
requires habitable terrestrial planets to remain strictly within the
boundaries of their habitable zone. The effects on our results of
relaxing this strict definition are also briefly discussed in \S5.

\subsection{Test Particle Approximation and Choice of Orbital Parameters}
\label{sub:tp}

To simplify and speed up the calculations, we make two important
approximations. First, we assume that terrestrial planets can be
treated as test particles. Second, we use the various system
parameters listed in Table~\ref{tab:one} and~\ref{tab:onebis} without
accounting for the observational errors associated with them.  We
justify these two approximations below.

Terrestrial planets with masses between $0.3 M_\oplus$ and $10
M_\oplus$ are typically $10^2$--$10^3$ times less massive than the
giant planets listed in Table~\ref{tab:one} and~\ref{tab:onebis}. The
large mass ratios suggest that a test particle approximation is
applicable. Since test particles are massless, they are
gravitationally perturbed by giant planets but do not perturb each
others nor the giant planet(s). This allows both the orbital
integration of 100 test particles simultaneously and the free removal
of any such particle at any time during integration (without change to
the system's Hamiltonian).  This is an obvious advantage for our
statistical study involving the orbital integration of 100 terrestrial
planets in the habitable zones of each of the $85$ systems of
interest.

Using the full potential of the test particle approximation by
integrating the orbital dynamics of each system with 100 terrestrial
planets at once also means that a unique set of values is used for the
stellar mass $M_*$,\footnote{The value of the stellar mass enters our
models in the determination of the radial extent of the habitable
zone.} the minimum planet mass $M_p \sin i$, the orbital period
$P_{\rm orb}$ and the eccentricity $e$. All these parameters are
subject to observational errors, however, and it is {\it a priori}
unknown what the influence of these errors might be on the resulting
statistics of dynamical habitability (i.e. the number of remaining
habitable planets).

To answer this question and validate the test particle approximation,
we carried out the following test. We selected three representative
single-planet systems of the dynamical classes II, III and IV, with
well-documented observational errors. We also attributed an arbitrary
$15\%$ error to the stellar masses. The list of planets, their orbital
parameters and associated errors for these three systems are given in
Table~\ref{tab:two}. For each system, we performed two sets of
integrations for a total of 100 terrestrial planets each. In the first
set of computationally-expensive calculations (experiment I), the
orbital dynamics of each of the 100 terrestrial planets is integrated
individually. The terrestrial planet semi-major axis is randomly
chosen within the habitable zone, with a random mass between 0.3 and
10~$M_\oplus$ and a random inclination relative to the giant planet's
orbit between 0 and $25^\circ$. The parameters for the giant planet's
orbit and the stellar mass (which determines the radial extent of the
habitable zone in each of the 100 independent integrations) were
generated as Gaussian deviates using the values and ($1 \sigma$)
errors quoted in Table~\ref{tab:two}. In the second set of
computationally-affordable calculations (experiment II), terrestrial
planets were treated as test particles (thus allowing the 100
terrestrial planet orbits to be integrated all at once), the giant
planet's orbital parameters and the stellar mass were fixed to their
nominal values and the same uniform distribution of inclinations
relative to the giant planet's orbit as in experiment~I was
assumed. In both experiments, terrestrial planets were initially put
on circular orbits and the giant planet's mass was set to its minimum
value (corresponding to $\sin i=1$ in column 3 of
Table~\ref{tab:two}), thus maximizing the effects of finite masses for
terrestrial planets in experiment I.

While experiment I represents a full statistical exploration of the
problem, experiment II is a simplified version which makes use of the
test particle approximation and neglects errors on estimated system
parameters. The two experiments were integrated for
$10^7$~yrs. Experiment I required approximately $\sim 6$ times more
CPU time than experiment II to complete and would thus be
prohibitively CPU--expensive to carry out for all extrasolar planetary
systems listed in Table~\ref{tab:one} and~\ref{tab:onebis}. Our test
shows, however, that experiment II provides results in reasonably good
agreement with those of experiment I for each of the three dynamical
classes.

\begin{figure}
\plotone{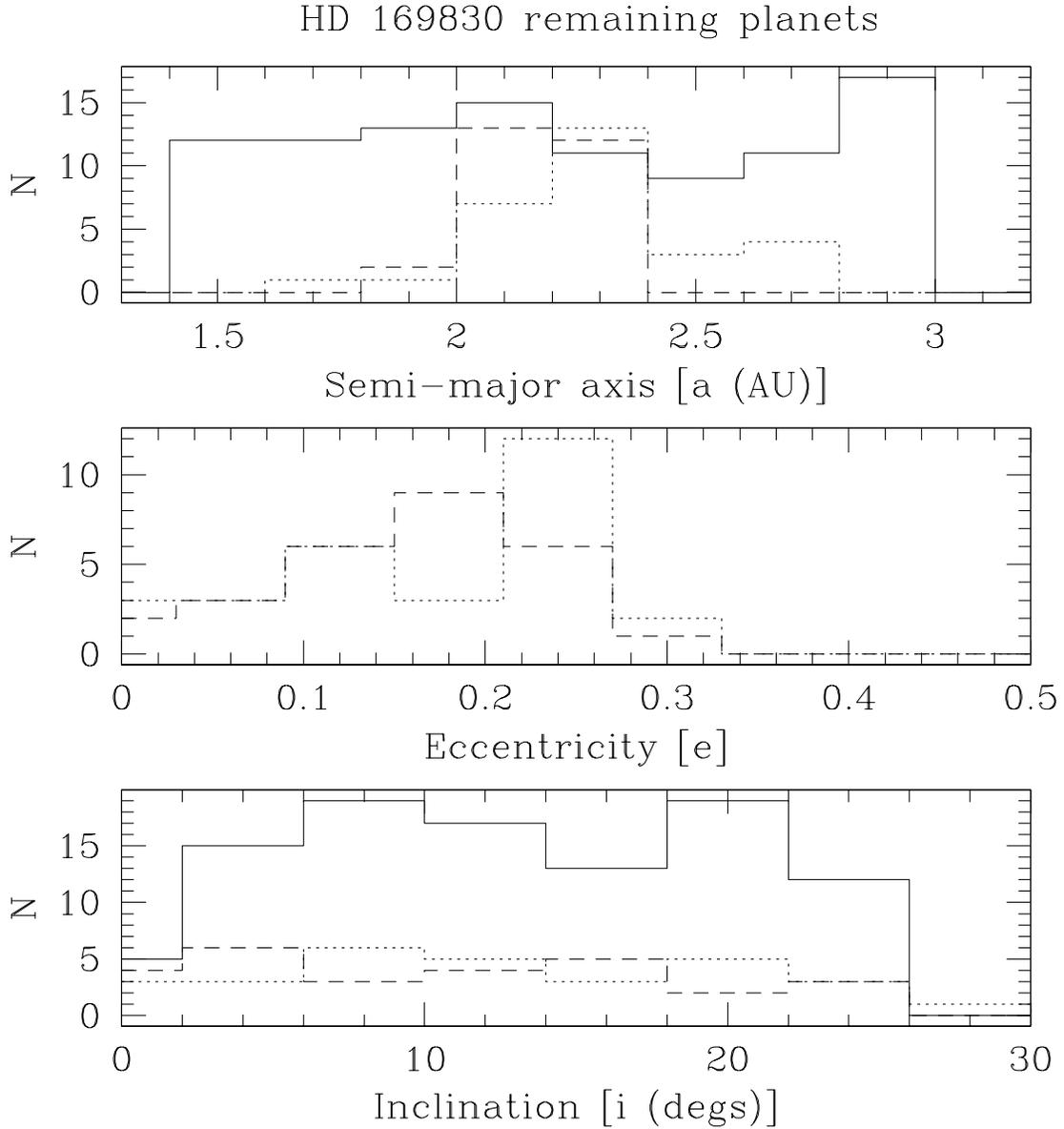}
\caption{Distribution of semi-major axes ($a$; upper panel),
eccentricities ($e$; middle panel) and inclinations ($i$; lower panel)
for terrestrial planets remaining within the habitable zone of the
system HD~169830 by the end of our integrations. The dotted line
corresponds to experiment I, while the dashed line corresponds to
experiment II. The mean and standard deviation of the distributions of
$a$, $e$ and $i$ in the two experiments agree well with each
other. The solid lines in the upper and lower panels show the initial
distributions of $a$ and $i$ in experiment II. The initial
eccentricity was zero for all terrestrial planets in these test
calculations. The habitable zone in HD~169830 extends from 1.4 to
3~AU.
\label{fig:one}}
\end{figure}

Figure~\ref{fig:one} shows the distributions of semi-major axes, $a$,
eccentricity, $e$, and inclination, $i$, at the end of experiments~I
(dotted line) and~II (dashed line) for planets that remained within
the habitable zone in the system HD~169830. The solid lines in the
upper and lower panels of figure~\ref{fig:one} show the initial
distributions of $a$ and $i$ in experiment II.  {The mean and standard
deviation of the final distributions of semi-major axis and
eccentricity in experiments I and II are: $a_{\rm I}=2.31 \pm 0.22$,
$a_{\rm II}=2.18 \pm 0.13$, $e_{\rm I}=0.17 \pm 0.09$ and $e_{\rm
II}=0.16 \pm 0.07$}. The good agreement between the final
distributions in experiments~I and~II supports the validity of the
approximations made in experiment~II in the present context of a
global statistical study. We note that the larger range of semi-major
axis, $a$, for remaining planets in experiment I (upper panel; dotted
line) is the result of variations of the stellar mass in that
experiment, which determines the radial extent of the habitable
zone. An additional discussion of the distributions shown in
figure~\ref{fig:one} can be found in \S4.1.

The statistics of the number of planets remaining in the habitable
zone(s) at the end of the numerical integrations (the quantity we are
most interested in) agrees reasonably well for each test system in
experiments~I and~II, as indicated by Table~\ref{tab:twobis}.  It is
reassuring to see the agreement even for the system HD~114783 which,
on the one hand, presents the largest fractional errors in system
parameters and, on the other hand, is probably the system most
sensitive to initial conditions among the three because of the partial
overlap between the giant planet's zone of influence and the habitable
zone. At the level of precision required by our study, these tests
successfully validate the test-particle approximation and show that
our results are not strongly affected by measurement errors on giant
planet orbital parameters (and stellar mass).

\subsection{Initial Conditions}
\label{sub:ic}

Besides the initial semi-major axis (randomly distributed within the
habitable zone), a number of orbital parameters for terrestrial
planets have to be chosen.  Angle variables, i.e. the argument of
pericenter, the longitude of ascending node and the mean anomaly, are
randomly distributed between $0^\circ$ and $360^{\circ}$ in all the
integrations for terrestrial planets. In systems with a single giant
planet, the giant is initially located at pericenter. In systems with
multiple giant planets, the observationally constrained values of the
time of periastron passage and argument of pericenter are consistently
used in the integrations. Terrestrial planets initial eccentricities
and inclinations (relative to the giant planet's orbit) must be
selected with care because the choice of distribution can potentially
affect the statistical results.

Initially circular orbits do not necessarily minimize resonant effects
with the giant planets. It is a well-known result of secular
perturbation theory that an orbit with finite eccentricity, $e$, and
inclination, $i$, can exhibit more stable properties than an orbit
with zero eccentricity and inclination (see, e.g., Murray \& Dermott
1999). Thus, by analogy with the observed distributions for the minor
bodies in the Solar System,\footnote{see {\tt
http://cfa-www.harvard.edu/iau/lists/MPDistribution.html}} we adopt
Rayleigh distributions for the initial eccentricities and inclinations
of terrestrial planets in our simulations, \be P(r) =
\frac{re^{-\frac{r^2}{2s^2}}}{s^2}, \;\;\; \mu=s\sqrt{\frac{\pi}{2}},
\;\;\; \sigma^2=\frac{4-\pi}{2}s^2, \ee where $r \in [0,\pi/2]$ for
the inclination and $r \in [0,1]$ for the eccentricity; $\mu$ and
$\sigma$ are the traditional mean and variance of the
distribution. Although each extra-solar system may have a specific
value of $s$ that maximizes stability for terrestrial planets in the
habitable zone, we use a unique value for all the systems to simplify
our statistical analysis. {We chose to fix the moments of the Rayleigh
distribution in all our integrations to $s(e)=0.05$ and $s(i)=5.2$ for
the initial eccentricity and inclination, respectively (unless noted
otherwise). This choice is consistent with the observed distributions
for minor bodies and typical values for terrestrial planets in the
Solar System.}  We have confirmed that, in some systems, starting all
terrestrial planets with zero eccentricity can result in a smaller
number of remaining habitable planets than when the above Rayleigh
distribution is used for initial eccentricities.

\begin{figure}
\plottwo{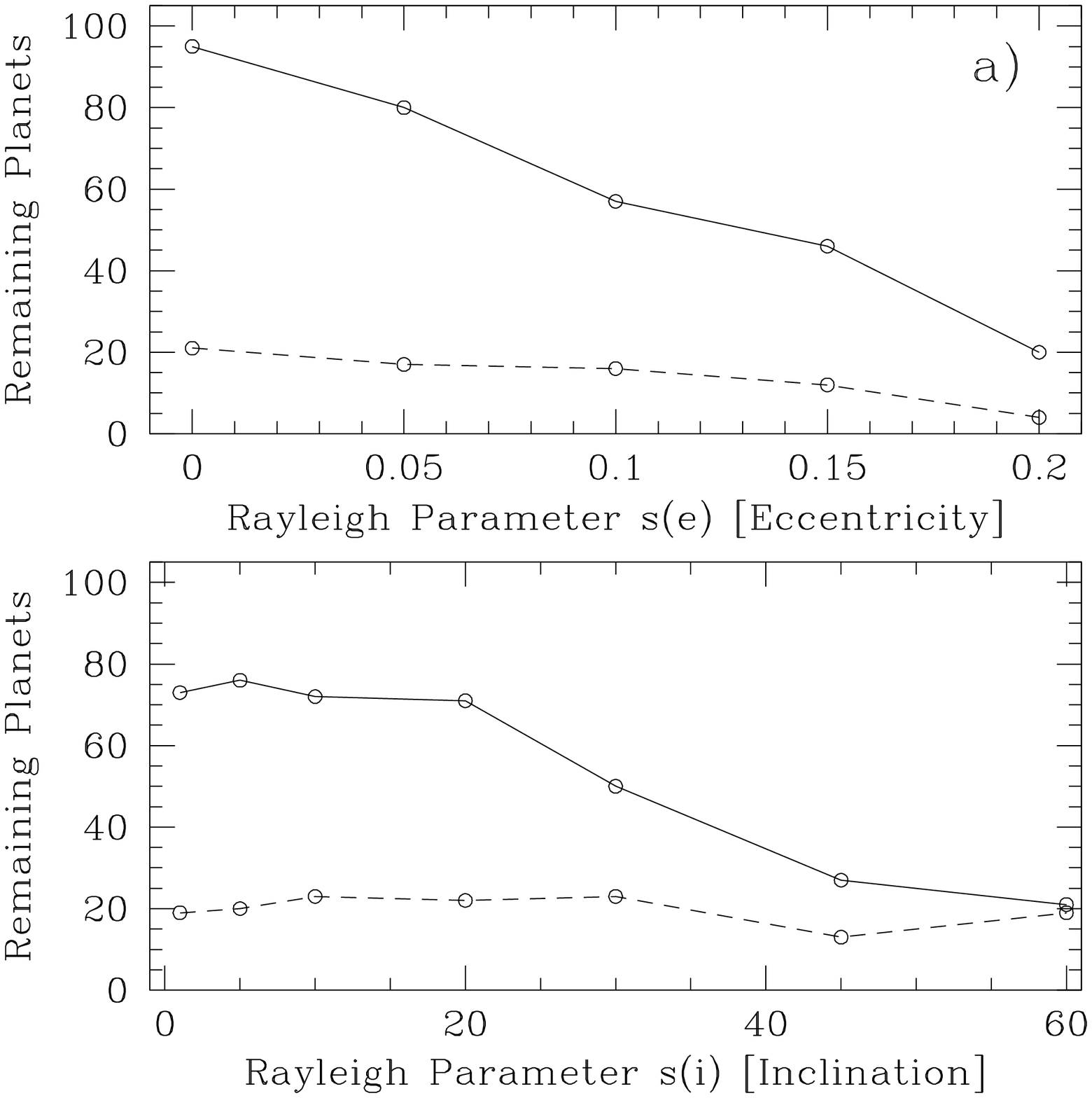}{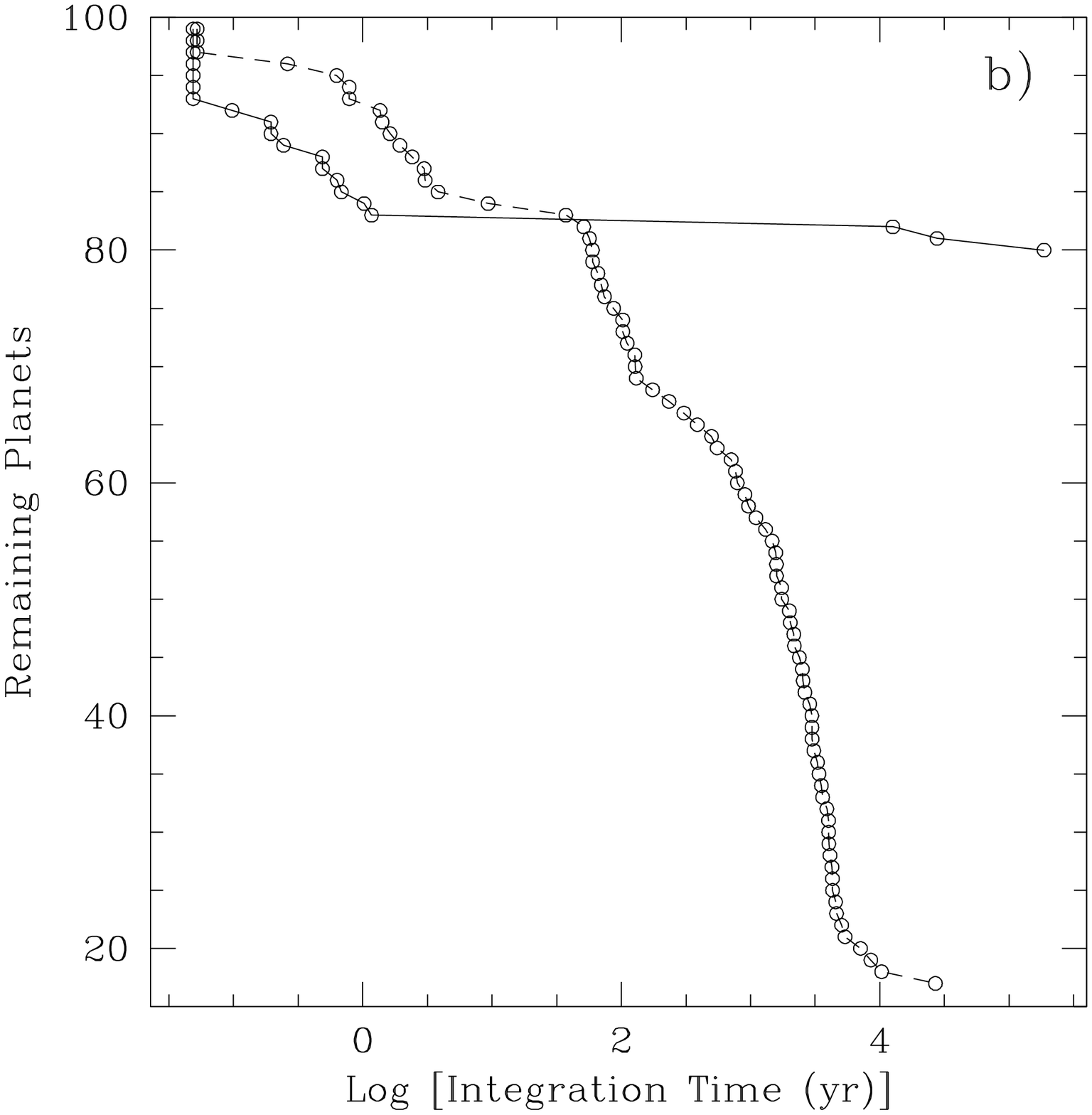}
\caption{(a) Number of remaining habitable terrestrial planets as a
function of the parameter $s$ determining the moments of their initial
Rayleigh distribution of eccentricity (upper panel) and inclination
(lower panel). In each panel, the solid line corresponds to the
Equivalent Solar System model while the dashed line correspond to the
HD~169830 model. (b) Evolution with time of the number of remaining
habitable planets in our fiducial models (with $s(e)=0.05$,
$s(i)=5.2$) of the Equivalent Solar System (solid) and HD~169830
(dashed).
\label{fig:one2}}
\end{figure}

We performed a number of numerical experiments, based on the
Equivalent Solar System defined below (see \S\ref{sec:equivss}), to
{confirm that this choice of $s(e)$ and $s(i)$ is reasonable.}
Figure~\ref{fig:one2}a shows how the number of remaining habitable
planets depends on the {values} of $s(e)$ and $s(i)$ for both the
Equivalent Solar System (solid line) and the test system HD~169830
(dashed line). The value of $s(i)$ was fixed at 5.2 when varying that
of $s(e)$ and the value of $s(e)$ was fixed at 0.05 when $s(i)$ was
varied. The typical evolution of the number of remaining habitable
terrestrial planets with integration time is shown for the two systems
in figure~\ref{fig:one2}b in models with $s(e)=0.05$ and $s(i)=5.2$.

The number of remaining habitable terrestrial planets is a decreasing
function of the value of $s(e)$, which determines both the mean and
standard deviation of the Rayleigh distribution of initial
eccentricities. More terrestrial planets cross the limits of their
habitable zone early in the simulation if they are given larger
initial eccentricities. The effect appears stronger for the Equivalent
Solar System only because in HD~169830 the giant planet is efficient
at perturbing terrestrial planet eccentricities even if they are
initially small, while Jupiter is much less efficient at this (see,
e.g., Fig.~\ref{fig:one4} and discussion below).

The results as a function of initial inclination do differ for the
Equivalent Solar System and HD~169830, however (lower panel;
Fig.~\ref{fig:one2}a). In the case of HD~169830, there is no obvious
dependence of the number of remaining habitable terrestrial planets
with the value of $s(i)$ (which determines both the mean and standard
deviation of the Rayleigh distribution of initial inclinations). For
$s(i) \gsim 20^\circ$, however, the number of remaining habitable
terrestrial planets for the Equivalent Solar System is reduced. This
is caused by the Kozai mechanism, which forces terrestrial planets to
experience periodic cycles with large maximum eccentricities provided
their initial inclination is in excess of $\sim 30-40^\circ$ (Kozai
1962). Those terrestrial planets starting with inclinations large
enough to be subject to the Kozai mechanism are eventually forced to
cross the limits of their habitable zone.

{The choice $s(i)=5.2$ guarantees that less than 0.001\% of our test
particules} start with inclinations in excess of $25^\circ$. We note
that this distribution of initial inclinations may imply that test
particules in some of the studied extrasolar planetary systems start
with inclinations in excess of the critical Kozai angle for that
system (since it is a function of the ratio of orbital distances
between the test particule and the giant; Kozai 1962). While this may
be the reason why some test particules leave their habitable zone in
some our simulations, we have not investigated this effect any
further.

\subsection{Giant Planet Masses and Inclinations}

The masses, $M_p$, and orbital inclinations, $i$, of the giant planets
listed in Table~\ref{tab:one} and~\ref{tab:onebis} are degenerate
because only minimum masses, $M_p \sin i$, are directly measured. In
general, no observational constraints exist on the inclination between
the orbital and sky planes. For randomly oriented systems and in the
absence of observational biases, the distribution of the {cosine of
the inclination ($\cos i$)} is uniform.  To sample the range of
possible inclinations (or equivalently giant planet masses), we
perform not just one but five integrations for each of the $85$
systems of interest, with $\sin i = 0.2 \cdot n + 0.1$; $n=0,...,4$.
{We also present our results averaged over these five inclination
values. While a uniform $\cos i$ distribution favors large $\sin i$
values, we have found little dependence of our results on inclination
except in a few isolated cases (see below).}  For simplicity, we have
assumed zero mutual inclinations between giant planets in
multiple-planet systems (there is support for low mutual inclinations
in at least one extrasolar planetary system; see Chiang, Tabachnik \&
Tremaine 2001). This assumption should not affect our results in any
significant way {(although it might affect the dynamics of a system
with eccentric giant planets close to resonance such as HD~82943)}.

We note that, for a deuterium-burning mass threshold of 13~$M_{\rm
Jup}$, the total sample of objects considered in our study (averaging
over the five inclination cases) comprises 78\% of giant planets and
22\% of brown dwarfs.  Since our analysis is dynamical in nature, we
see no reason not to include brown dwarfs in our calculations.
Heating of terrestrial planets by brown dwarfs is negligible at the
typical distances considered, so that we neglect this effect on the
extent of the habitable zone (giant planet or brown dwarf heating is
also negligible when considering the habitability of moons; Williams
et al. 1997). The subdivision by dynamical class of the total sample
of objects (averaged over inclinations) is as follows: 26\% of
class--I objects, 13\% of class--II objects, 26\% of class--III
objects and 35\% of class--IV objects.

\subsection{Equivalent Solar System}
\label{sec:equivss}

Our definition of dynamical habitability is arbitrary. Although it is
based on the intuitive idea that the gravitational influence of a
giant planet might prevent the presence of terrestrial planets within
the habitable zone of a planetary system, the quantitative measure of
dynamical habitability that we obtain from our simulations (i.e. the
number of remaining habitable planets after a fixed amount of
integration time) is difficult to interpret by itself. It is possible,
however, to compare this measure with its equivalent for the Solar
System, in which we know of the presence of an habitable and inhabited
planet, the Earth.

To make this comparison, we need to define an ``Equivalent Solar
System'', which is the Solar System as we would know it had it been
detected only from tens of parsecs away with current radial velocity
surveys. Our motivation to use this Equivalent Solar System comes from
the possibility that additional, undetected planets in any of the
extrasolar planetary systems studied here may modify their dynamical
habitability properties significantly. Therefore, our Equivalent Solar
System is defined as the Solar System subject to the same
observational selection effects as extrasolar planetary systems. It
only contains Jupiter, since this is the only planet that would have
been detected by current surveys. {An example of the biases avoided by
comparing extrasolar planetary systems to the Equivalent Solar System
is the destabilizing effect of overlapping resonances caused by
Jupiter and Saturn's orbital precessions in the Asteroid Belt: this
effect is absent from the Equivalent Solar System model in the same
way the influence of yet undetected giant planets in our models of
extrasolar planetary systems is ignored.}

The habitable zone of our Equivalent Solar System is taken to extend
from 0.7 to 1.3~AU, for consistency with the values listed in
Table~\ref{tab:one} and~\ref{tab:onebis}. The zone of influence of
Jupiter extends down to 3.9~AU only, making the Equivalent Solar
System a class II system according to our definition in
\S\ref{sec:class}. By analogy with what is done for extrasolar
planetary systems, we also investigate the effect of varying the mass
of the giant planet in the Equivalent Solar System by running five
different models with an assumed $M_p \sin i =M_{\rm jup}/2$ and $\sin
i = 0.2 \cdot n + 0.1$; $n=0,...,4$ (the middle case $\sin i =0.5$
thus corresponds to $M_p =M_{\rm jup}$ and other cases show what the
effects of varying Jupiter's mass are).

\section{Results}

We have integrated the orbital dynamics of all $85$ extrasolar
planetary systems listed in Table~\ref{tab:one} and~\ref{tab:onebis}
and that of the Equivalent Solar System defined above, with 100
terrestrial planets (treated as test particles) in their habitable
zones, for $10^6$~years. We describe our results in the following
subsections.

\subsection{Terrestrial Planet Orbits}

\begin{figure}
\plotone{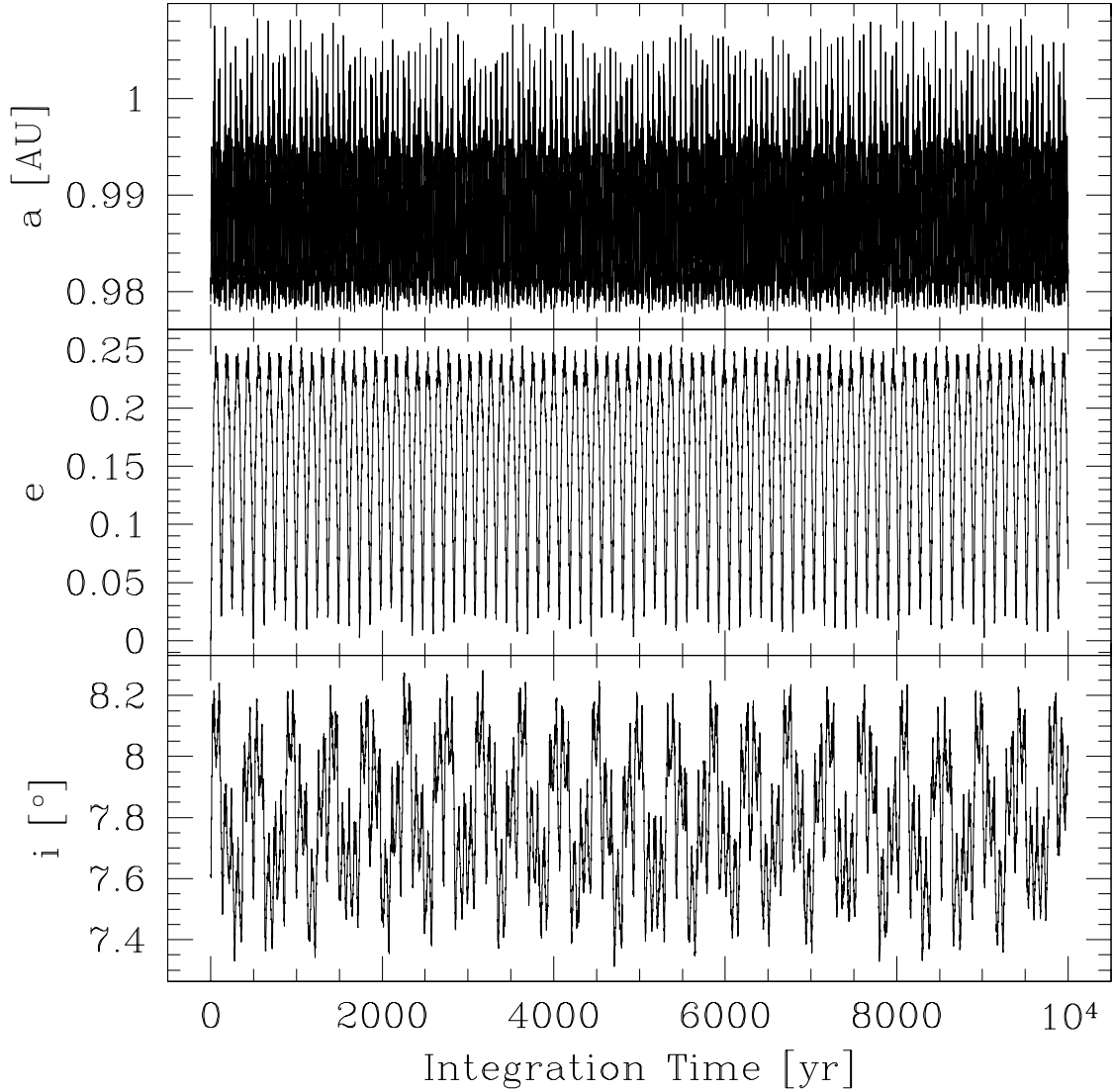}
\caption{Example of the evolution of the orbital elements $a$
(semi-major axis), $e$ (eccentricity) and $i$ (inclination) of a
terrestrial planet remaining within the habitable zone in our test
model of the system HD~114783.
\label{fig:one3a}}
\end{figure}

\begin{figure}
\plotone{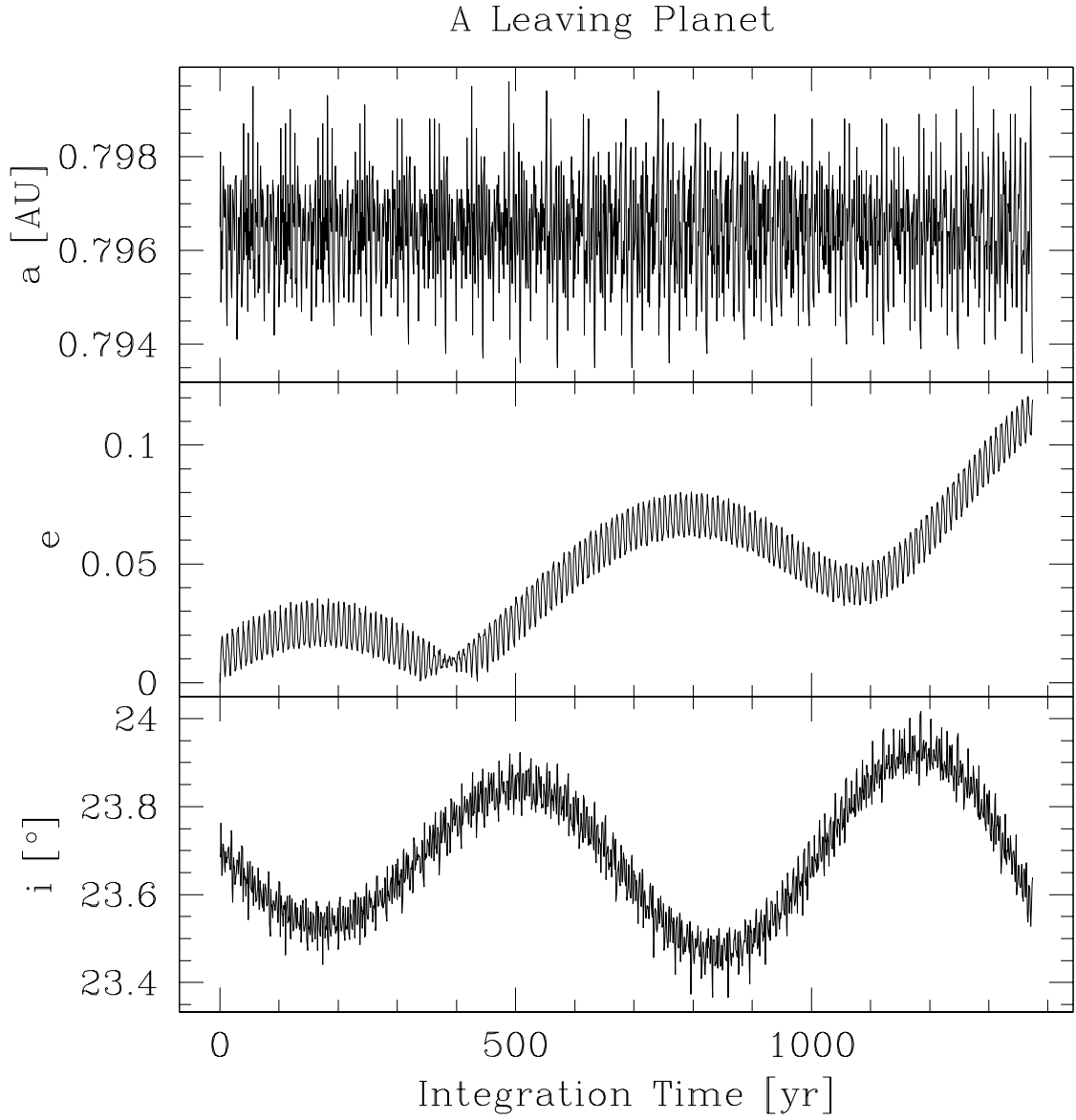}
\caption{Same as Fig.~\ref{fig:one3a} for a planet leaving the inner
edge of the habitable zone before $10^6$~years of integration.
\label{fig:one3b}}
\end{figure}

\begin{figure}
\plotone{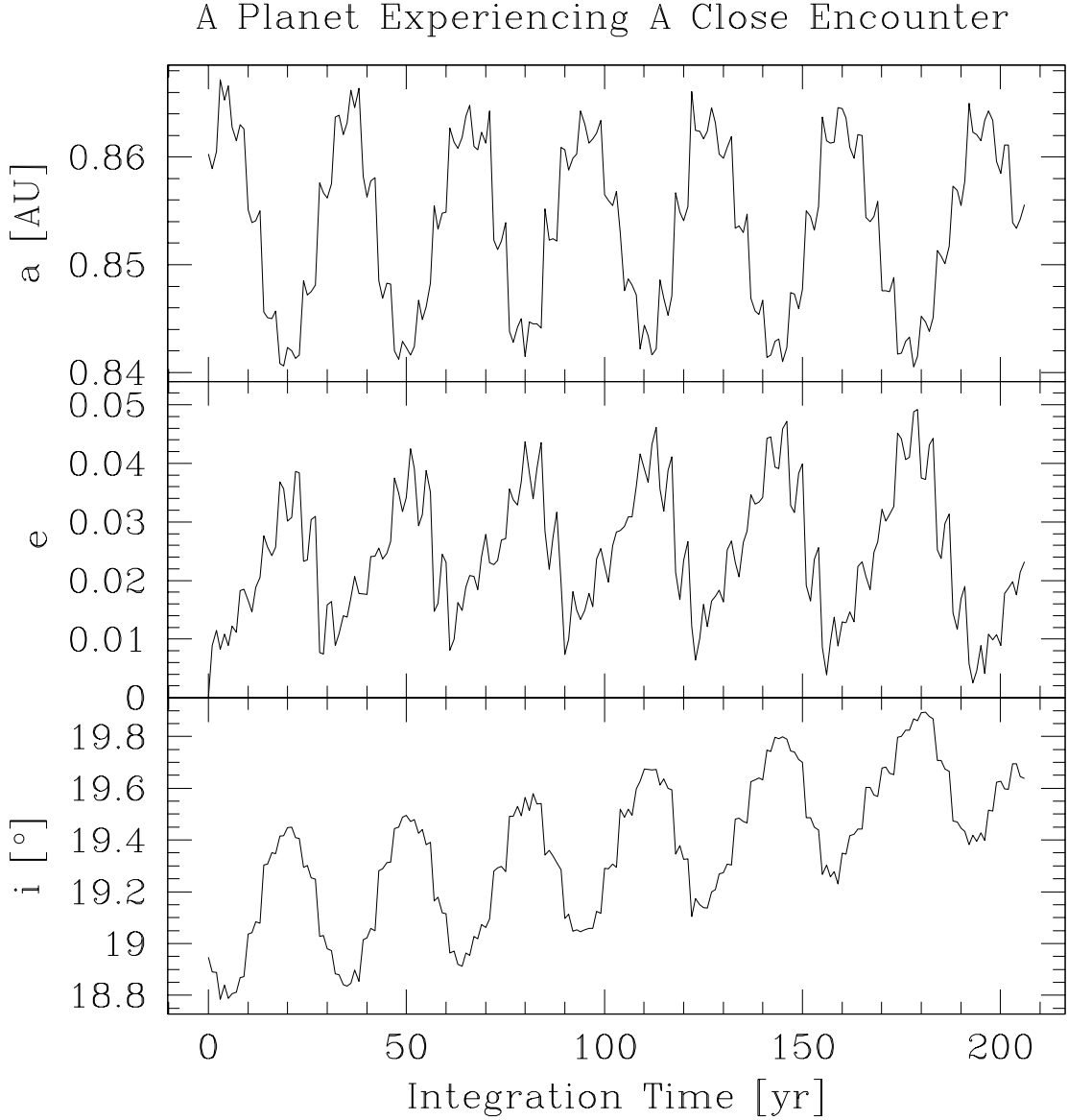}
\caption{Same as Fig.~\ref{fig:one3b} for a planet experiencing a
close encounter with the giant planet. While the terrestrial planet is
located within the giant planet's zone of influence as we defined it
(with an inner edge at $0.83$~AU), it takes some time before the
terrestrial planet actually penetrates the 3--Hill--radii sphere
around the giant planet.
\label{fig:one3c}}
\end{figure}

Figure~\ref{fig:one3a}--\ref{fig:one3c} illustrate the 3 types of
terrestrial planet orbital evolution seen in our integrations through
examples taken from our test particle (experiment II) integration of
the HD~114783 system (\S\ref{sub:tp}). Some terrestrial planets
experience a close encounter with a giant planet, generally early
during the simulation (Fig.~\ref{fig:one3c}). Others remain at all
time habitable, with a finite eccentricity allowing them, given their
semi-major axis, not to cross the limits of their habitable zone
(Fig.~\ref{fig:one3a}). Others see their eccentricity experience
important secular variations, until they first cross the limits of the
habitable zone (at which point they are no longer considered
habitable; Fig.~\ref{fig:one3b}).

\begin{figure}
\plotone{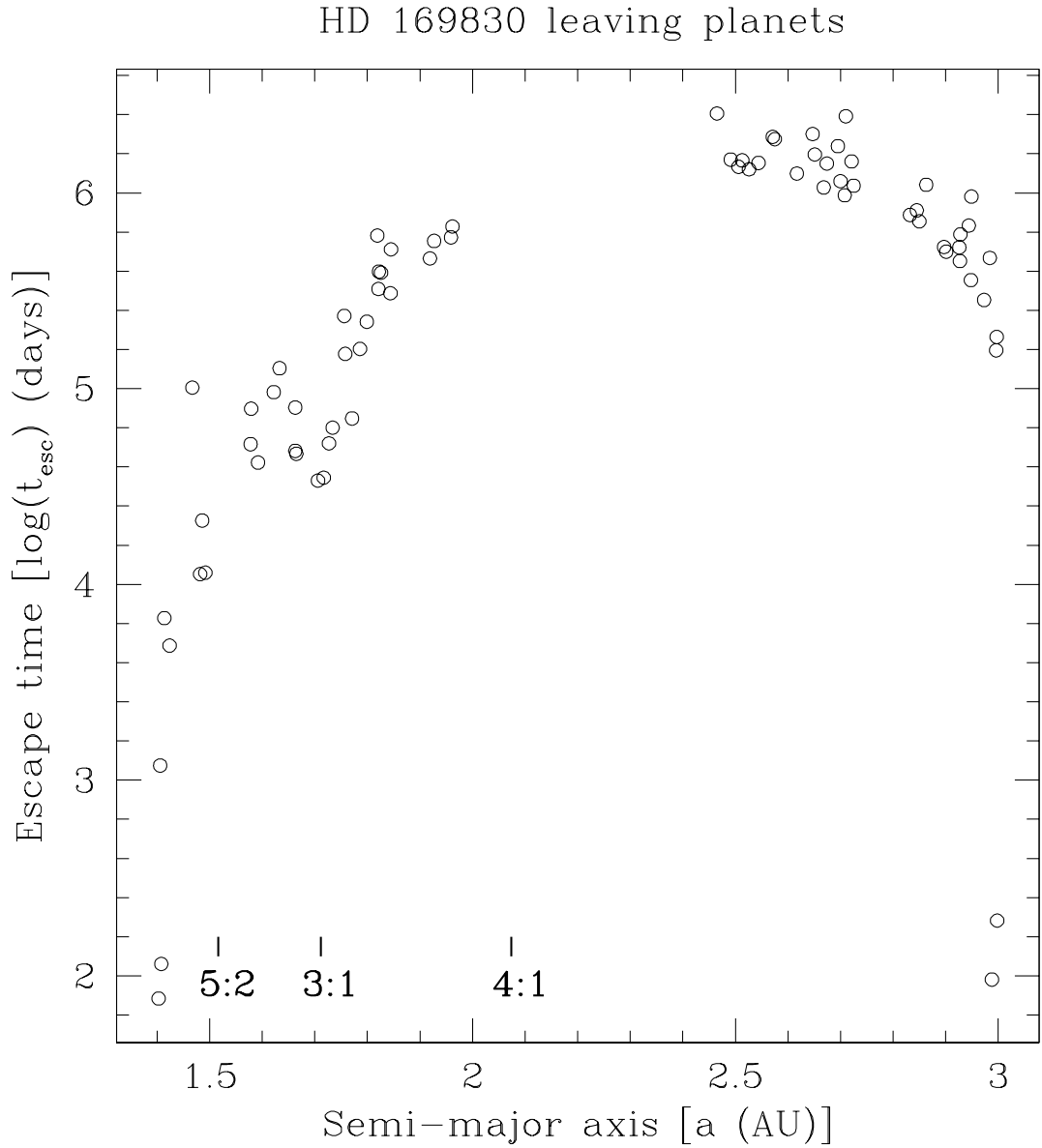}
\caption{Time of {escape} out of the habitable zone versus semi-major
axis at that time for the 71 leaving planets in the HD~169830 system
(test experiment II). The locations of strong mean-motion resonances
of highest order in the habitable zone are also indicated.
\label{fig:two}}
\end{figure}

\begin{figure}
\plotone{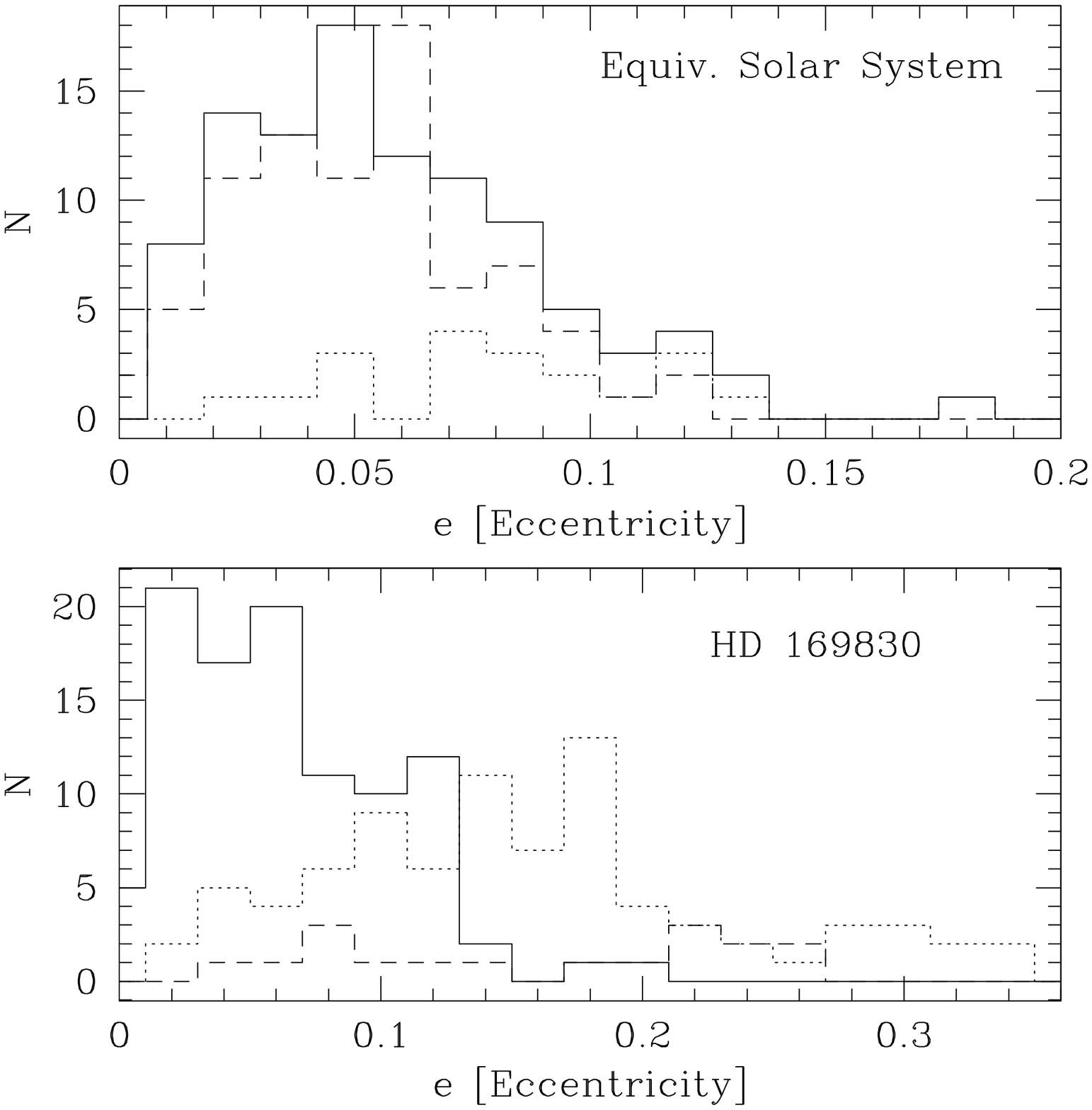}
\caption{Shows the evolution of the eccentricity distributions of
terrestrial planets in the Equivalent Solar System (upper panel) and
HD~169830 (lower panel) models. In each panel, the initial (Rayleigh)
distribution is shown as a solid line, the final distribution for
remaining habitable planets is shown as a dashed line and the
distribution for planets leaving the habitable zone (at the time they
first leave) is shown as a dotted line.
\label{fig:one4}}
\end{figure}

We found that secular eccentricity variations {(see, e.g., Murray \&
Dermott 1999)} play an important role for dynamical habitability.  As
the eccentricity of terrestrial planets are subject to these
variations, one expects those closest to the inner and outer edges of
the habitable zone to be the first to cross its limits. This { gradual
escape process} is shown in figure~\ref{fig:two}, where the time at
which planets first leave the habitable zone is plotted as a function
of their semi-major axis at that time (for the experiment--II test
calculation on HD~169830). Note the asymmetry in the {escape} process
which makes planets closer to the inner edge of the habitable zone
leave its boundaries faster than those close to the outer edge. We
interpret this as being due to the presence of the giant planet {on
the inner side} of the habitable zone in this system, making its
influence on planets in the inner region of the habitable zone
stronger. The effect of the 3:1 mean-motion resonance is also visible
in figure~\ref{fig:two}, while the 5:2 and 4:1 resonances seem to have
weaker effects on the {escape} process.

In general, mean-motion resonances are not the dominant effect leading
to {the escape of terrestrial planets from} their habitable zone in
our simulations. Table~\ref{tab:threebis} lists $p/(p+q)$ mean-motion
resonances of order $q=1,2,3$ up to $p=10$ in the habitable zones of
class~I and class--II systems only. None of the class--I systems has
such a strong resonance within its habitable zone, while many of the
class--II systems do. Figure~\ref{fig:two} shows, however, that the
effect of such resonances can be localized and that instead most {
escapes occur} because of secular eccentricity variations.  We also
expect mean-motion resonances to be present in the habitable zones of
class--III and class--IV systems, but many terrestrial planets also
experience close encounters with a giant planet in these systems. We
have not studied the relative importance of mean-motion resonances and
close encounters in class--III and class--IV systems in more detail.

The asymmetry of the {escape} process in figure~\ref{fig:two} suggests
that secular eccentricity variations should be reduced in systems
where the giant planet is further away from the habitable zone. This
is confirmed by Figure~\ref{fig:one4}, which compares the initial and
final distributions of eccentricity in our fiducial models of the
Equivalent Solar System and HD~169830. While the eccentricity
distribution of terrestrial planets in the Equivalent Solar System
model has barely evolved (whether these planets remained habitable or
crossed the limits of their habitable zone), the evolution, up to
eccentricities $\sim 0.3$ for remaining habitable planets, is obvious
in the model of HD~169830.

The relative variation of insolation flux for a planet with
eccentricity $e$ is the square of the ratio of apoastron to periastron
distance, which amounts to $(1+e)^2/(1-e)^2$. While global variations
of order $20-30\%$ of the insolation flux on the Earth did occur in
the past when its eccentricity was as much as $\sim 0.05$ (standard
Milankovitch theory\footnote{see, e.g., {Hays, Imbrie \& Shackleton
1976;} {\tt
http://earthobservatory.nasa.gov/Library/Giants/Milankovitch/}}), the
expected variations for some of the more eccentric habitable planets
found in our simulations would be much larger than this (e.g. $\sim
125\%$ for $e \sim 0.2$). It {is unclear whether planets experiencing
such strong seasonal climate variations provide a good environment for
the development of life (see Williams \& Kasting 1997 for a related
discussion of the climatic effects of {\it regional} flux variations
caused by obliquity changes).}

\subsection{Global Statistics for Dynamical Habitability}

Table~\ref{tab:three} shows results from our large set of numerical
simulations, in terms of the number of terrestrial planets remaining
within their system's habitable zone after $10^6$~years of
integration. The number of remaining habitable planets is given for
each of the five different orbital inclinations considered, together
with the average over these five cases for each of the systems of
interest. For comparison, the number of remaining habitable planets
after $10^6$~years in models of the Equivalent Solar System are also
listed.\footnote{In all the models listed in Table~\ref{tab:three},
the adopted Rayleigh distribution of initial eccentricities causes a
moderate number of terrestrial planets to cross the limits of their
habitable zone during their first orbital revolution.} The mean and
standard deviation of the final distributions of semi-major axis
($a$), eccentricity ($e$) and inclination ($i$) of remaining habitable
planets are also given in the last three columns of
Table~\ref{tab:three}, individually for each extrasolar planetary
system.

It is clear from Table~\ref{tab:three} that our results show two
populations of systems with drastically different dynamical
habitability properties. About $1/4$ of all systems retain a high
percentage ($\sim 60$--$80\%$) of their initially habitable
terrestrial planets, as does the Equivalent Solar System. This group,
which are comparable to our Solar System in terms of dynamical
habitability, is mostly comprised of systems containing close-in giant
planets. The proximity of the giant planets from the parent star in
these systems, and their predominantly small eccentricities, prevent
them from gravitationally interacting much with terrestrial planets in
the habitable zone. Of the 17 systems closer than 20~pc from us
(probably the maximum distance of potential targets for {\it TPF}),
only $\sim 1/3$ retain a high percentage of their initially habitable
terrestrial planets and are thus comparable to the Solar System in
terms of dynamical habitability.

A large number of systems (more than $1/2$) were not able to retain
any (or almost any) terrestrial planet confined to their habitable
zone by the end of the simulations. This group is mostly comprised of
systems containing more distant giant planets on orbits with
substantial eccentricities. Both the location of habitable zones
typically around $0.5$--$2.0$~AU and the large effective zone of
influence of giant planets in these systems (because of their
substantial eccentricities) are responsible for the strong perturbing
influences on habitable terrestrial planets in these cases. The
generally larger masses of distant extrasolar giant planets (as
compared to generally less massive close-in planets; see
Table~\ref{tab:one} and Tabachnik \& Tremaine 2002) also contribute to
their stronger perturbing influence.

There is also a number of intermediate systems retaining a small ($\ll
80\%$) but finite number of habitable terrestrial planets. HD~169830
is an example of such systems where the proximity of the giant planet
from the habitable zone leads to efficient {escape} of initially
habitable planets (see discussion in \S4.1). In these intermediate
systems, we expect dynamically habitable planets to be rather
localized within their system's habitable zone: in the central regions
of the habitable zone in class--II systems (see example of HD~169830
in Fig.\ref{fig:one}) and on the opposite side of the habitable region
where close encounters with the giant planet occur in class--III
systems. It is interesting to note that multiple-planet systems, as a
group, are less dynamically habitable than single-planet systems. This
reflects the fact that most multiple-planet systems (with the possible
exceptions of 47~UMa and 55~Cnc) possess at least one class--III or
class--IV planet whose influence is very disturbing to dynamical
habitability. We have also observed that for certain choices of $\sin
i$, some of the multiple giant planet systems themselves become
unstable.

Table~\ref{tab:four} shows the statistics of remaining habitable
planets grouped by dynamical class. As expected, the stronger the
overlap between the giant planet's zone of influence and the habitable
zone, the smaller is the number of remaining habitable terrestrial
planets. Class I giant planets, as a group, are as dynamically
habitable as our Solar System. The dynamical habitability of class--II
and class--III systems is less than what we could have naively
expected, however. While in class--II systems there is no overlap
between the giant planet's zone of influence and the habitable zone,
only $\sim 1/3$ of initially habitable terrestrial planets for this
group remain confined to the habitable zone after $10^6$~years, on
average (see Table~\ref{tab:four}). This simply indicates that our
definition of the zone of influence does not capture the effect of
secular eccentricity variations caused by giant planets on the
dynamical habitability of terrestrial planets. Similarly, the number
of terrestrial planets remaining strictly confined to their habitable
zone is almost zero for class III systems, as a group, while the
overlap between the defined zone of influence and the habitable zone
is only partial in their case.

Table~\ref{tab:four} also lists for each dynamical class the average
percentage of terrestrial planets that leave their habitable zone
because of a close encounter with a giant planet. While close
encounters never occur in class--I and class--II systems, they play an
important role for the dynamical habitability of class--III and
class--IV systems (by definition).  The last four columns of
Table~\ref{tab:four} show, integrated over all members of each
dynamical class (class IV excluded), the total number of remaining
habitable planets and the mean and standard deviation of their final
distributions of semi-major axis ($a$), eccentricity ($e$) and
inclination ($i$).  Terrestrial planet inclinations do not vary much
in our integrations, so that the mean inclination for remaining
habitable planets is close to what the initial mean was ($\simeq
6.3^\circ$). The mean eccentricity of remaining habitable terrestrial
planets in class--I systems is also not that different from its
initial value because of the only moderate influence of the giant
planet on habitable planets in these systems. On the other hand, the
mean eccentricity of remaining habitable terrestrial planets in
class--II and class--III systems is significantly larger than its
initial value because the perturbing influence of giant planets on the
habitable zone region is felt more strongly in many such systems (but
not in the Equivalent Solar System).

\section{Discussion}

We have presented an extensive set of dynamical integrations of known
extrasolar planetary systems to evaluate how likely they are to harbor
habitable terrestrial planets. A number of choices made in our
numerical simulations, such as the initial conditions for terrestrial
planet integrations, are arbitrary. In making these choices, however,
we generally tried to pick conditions that maximize dynamical
habitability. For example, by adopting the test particle
approximation, we implicitly rejected the possibility that more than
one ``interacting'' terrestrial planet could be present in the
habitable zone of a system, which would generally reduce mutual
chances of survival in that zone.

Despite this arbitrariness, we believe that our results offer a
measure of dynamical habitability in extrasolar planetary systems
through the important comparison with the Equivalent Solar System. For
example, choosing the initial eccentricity of all our terrestrial
planets to be zero instead of the Rayleigh distribution described in
\S\ref{sub:ic} would somewhat affect the number of remaining habitable
planets reported in Table~\ref{tab:three} and~\ref{tab:four} (see
Fig.~\ref{fig:one2}). However, it would also affect the results for
the Equivalent Solar System so that the comparison remains
meaningful. Similarly, while more terrestrial planets may be found to
leave their habitable zone after $10^7$~yrs than after $10^6$~yrs of
integration (for instance), our practical choice of $10^6$~yrs equally
affects results for the Equivalent Solar System and all extrasolar
planetary systems.

Our results on the dynamical habitability of extrasolar planetary
systems are affected by the ignorance resulting from observational
selection effects. In several cases, multiple--planet systems were
first discovered through the influence of the one planet closest to
the parent star and additional planets were only subsequently
detected. This bias affects all known extrasolar planetary systems. As
additional planets are detected in a system, we expect their
gravitational influence to make the system equivalent or less
dynamically habitable than found in our models.\footnote{Stabilizing
secular resonances involving the presence of additional planets are
possible, but they may be exceptions.} Thus, as we learn more in the
future about the planetary content of already known extrasolar
planetary systems, their dynamical habitability (as estimated in our
models) may well diminish. The number of remaining habitable planets
listed in Table~\ref{tab:three} may thus correspond to an upper limit
to the dynamical habitability of these systems.

In that respect, it is significant that most of the extrasolar giant
planets recently discovered (compare Table~\ref{tab:onebis} to
Table~\ref{tab:one}) are of the distant, eccentric type which is most
disturbing to potentially habitable terrestrial planets. The
discoveries of 55~Cnc~d and Gl~777A~b show, however, that Doppler
surveys just began to reveal distant, low-eccentricity giant planets
which are more akin to the Solar System giants and do not perturb
potentially habitable terrestrial planets significantly. While the
Solar System is expected to remain roughly as dynamically habitable as
estimated in our equivalent--model when the other three giant planets
are included, the dynamical habitability of known extrasolar planetary
systems will mostly depend on whether radial velocity surveys
establish or rule out the presence of giant planets with strongly
perturbing influences at $\sim 0.5$--$2$~AU distances from the parent
star.

Our results are generally in good agreement with those of previous,
system--specific studies, when one accounts for the discovery of
additional planets in some of the systems of interest (e.g. in 47 UMa;
Gehman et al. 1996; Rivera \& Lissauer 2000; 2001). The work closest
to ours is that of Jones et al. (2001), where the authors integrated
the orbital dynamics of potentially habitable terrestrial planets in
only four systems, but for a much longer period of time than in our
simulations. While our results are consistent with those of Jones et
al. for the most part, our investigation also differs in a number of
ways, most importantly in our emphasis on a statistical comparison to
the Equivalent Solar System and in the conservative definition we
adopted for the dynamical habitability of terrestrial planets (Jones
et al. only require the planet {\it semi-major axis} to remain at all
time within the habitable zone).

The issue of the orbital requirements for a terrestrial planet to be
dynamically habitable is important. While we argued that even
temporary excursions outside of the habitable zone may be damaging to
the habitability of a terrestrial planet, based on the relatively
short radiative time constant of the Earth's atmosphere (\S3.1), one
could use the much larger thermal inertia of Venus' dense atmosphere
as a possible counter-example. It will thus be important to determine
more accurately in the future what the orbital requirements for
dynamical habitability are. We note that even if we were to relax our
definition of dynamical habitability by only requiring the terrestrial
planet's semi-major axis to remain confined to the habitable zone, our
main conclusions would not be much affected. Close encounters would
still efficiently remove terrestrial planets from the habitable zones
of all class--IV systems (i.e. systems containing at least one
class--IV planet), which constitute about half the total sample. The
number of remaining habitable planets in intermediate class--II and,
to some extent, class--III systems would be significantly larger than
estimated in our study because these are the systems where secular
eccentricity variations play the most important role. It is also
possible that mean-motion resonances become more determinant for the
dynamical habitability of class-III systems in that case.

It is interesting to speculate on the possible existence of planetary
systems containing only terrestrial planets (i.e. lacking gaseous
giant planets). The observed distribution of circumstellar disk masses
around young stars suggests that this situation could occur in systems
with low-mass disks (see, e.g., Beckwith \& Sargent 1996). A possible
measure of the expected number of systems containing only terrestrial
planets comes from the inference by Tabachnik \& Tremaine (2002),
based on the observed planetary mass function, that $\sim 3 \%$ of
Sun-like stars harbor planets in the gaseous giant mass range, while a
larger $\sim 18 \%$ may harbor terrestrial planets.

Systems containing only terrestrial planets, if they exist, would not
be subject to the perturbing influence of giant planets which was the
focus of the present study.  From that point of view, they would
appear as better environments for habitable planets than currently
known extrasolar systems. Even this point is not clear, however,
because another dynamical effect may reduce the habitability of
terrestrial planets in the habitable zones of these hypothetical
systems: the standard scenario for the formation of a cometary Oort
cloud in our Solar System (see, e.g., Duncan, Quinn \& Tremaine 1987)
would not apply to these systems, so that the number of remaining
minor bodies in the inner regions of these planetary systems may be
large. The resulting secular flux of asteroid impacts on terrestrial
planets may then be large enough to have damaging effects on
potentially habitable planets.

Interestingly, this additional dynamical effect, which deserves
further attention, may also be important for a large number of the
known extrasolar planetary systems that were classified as being as
dynamically habitable as the Solar System in our study. Most of those
possess close-in giant planets (located further in than the habitable
zone) which would not protect in any way habitable terrestrial planets
from a large flux of cometary impacts (as does Jupiter in the Solar
System). It is unclear how strong cometary fluxes in these systems
should be because giant planet migration may have affected the
properties of their cometary clouds significantly (Hansen
2002). Finally, we note that the planetisimal-scattering migration
scenario may not allow for the formation of any terrestrial planet in
systems containing close-in giant planets because, in this picture,
migration from large distances is associated with the efficient
clearing of the pre-existing disk of planetisimals (Murray et
al. 1998).

The discovery of distant extrasolar giant planets with orbital
properties preventing them from perturbing potentially habitable
terrestrial planets and such that they offer a protection from
possibly large fluxes of cometary impacts may thus be of great
interest in the future. Marcy et al. (2002), based on their discovery
of 55 Cnc~d, estimate that giant planets with orbital characteristics
comparable to those of Jupiter exist around about 1 out of 50 nearby
Sun-like stars. Microlensing searches for planets in the 1--10~AU
range could ideally complement radial velocity surveys for this task
(see, e.g., Gaudi 2002), since the latter become increasingly
time-prohibitive at large orbital distances.

\section{Conclusion}

We investigated the orbital dynamics of known extrasolar planetary
systems in the presence of potentially habitable terrestrial
planets. We adopted a conservative definition of dynamical
habitability by requiring that habitable planets never cross the
limits of their habitable zone. We find that more than half the known
extrasolar planetary systems are unlikely to harbor habitable
terrestrial planets because of the strong perturbing influence of
giant planet(s) in the vicinity of their habitable zones. Still, about
a quarter of the known systems, including a third of the potential
(nearby) targets for {\it TPF}, are promising in that they appear as
dynamically habitable as our own Solar System. Some terrestrial
planets labeled as dynamically habitable in our study have substantial
eccentricities ($e > 0.1$) which may lead to large seasonal climate
variations and potentially damaging effects for habitability.

\section*{Acknowledgments}

The authors are grateful to B. Hansen, J. Lissauer and S. Tremaine for
useful discussions and to L. Hernquist for comments on the manuscript.
Support for this work was provided by NASA through Chandra Fellowship
grant PF9-10006 awarded by the Smithsonian Astrophysical Observatory
for NASA under contract NAS8-39073. ST acknowledges support from an
ESA Research Fellowship.

\clearpage

\begin{table*}
{\tiny
\caption{74 EXTRASOLAR GIANT PLANETS}
\begin{center}
\begin{tabular}{lccccrcccc} \hline \hline
\\
Planet & Distance& $M_*$ &$M_p \sin i$ & $a$  & $P_{\rm orb}$ & $e$ & Habitable Zone &  Zone of Influence & Class\\
& (pc) & ($M_\odot$) & ($M_{\rm Jup}$) & (AU) & (days) & & (AU) & (AU)\\
(1)&(2)&(3)&(4)&(5)&(6)&(7)&(8)&(9)&(10)\\
\\
\hline
\\
HD 83443 b  &      43.54 &   0.79  &   0.350 &   0.0380  &   2.9853  &       0.0800&   0.50/1.00    &    0.027/0.049  &    I\\
HD 46375 b   &      33.40 &    1.00 &    0.249 &   0.0410  &   3.0240     &    0.0000 &  0.70/1.30  &      0.034/0.048 &     I\\
HD 187123 b     &   50.00  &   1.06  &   0.520 &   0.0420 &    3.0900    &     0.0300  & 0.75/1.40  &      0.032/0.052  &    I\\
HD 179949 b   &    27.00 &    1.24 &    0.840&    0.0450  &   3.0930     &    0.0500 &  1.10/2.25   &     0.033/0.057  &    I\\
Tau Boo b     &    15.00  &   1.30   &  4.090  &  0.0500   &  3.3120  &       0.0000  & 1.25/2.70  &      0.031/0.069  &    I\\
BD -10316 b   &     ?   &     1.10  &   0.480 &   0.0460  &   3.4870   &      0.0000  & 0.85/1.60  &      0.037/0.055  &    I\\
HD 75289 b     &    28.94 &   1.05  &   0.420 &   0.0460  &   3.5100    &     0.0540  & 0.75/1.40  &      0.035/0.057   &   I\\
HD 209458 b    &    47.00  &   1.05  &   0.690 &   0.0450  &   3.5247  &       0.0000  & 0.75/1.40  &      0.035/0.055   &   I\\
51 Peg b    &       14.70  &   0.95  &   0.440 &   0.0512  &   4.2300   &      0.0130  & 0.70/1.30    &    0.040/0.062   &   I\\
Ups And b     &    16.50  &   1.30 &    0.710 &   0.0590  &   4.6170  &       0.0340 &  1.25/2.70     &   0.045/0.073  &    I(IV)\\
HD 68988 b    &    58.00   &  1.20   &  1.900  &  0.0710   &  6.2760   &      0.1400  & 0.93/1.80     &   0.040/0.102  &    I\\
HD 168746 b   &     43.12 &   0.92   &  0.240  &  0.0660 &    6.4090   &      0.0000 &  0.65/1.25     &   0.055/0.077  &    I\\
HD 217107 b   &    37.00 &    0.98   &  1.280 &   0.0700  &   7.1270     &    0.1400 &  0.70/1.30 &       0.040/0.100  &    I\\
HD 130322 b   &    30.00  &   0.79  &   1.080 &   0.0880  &  10.7240    &     0.0480  & 0.50/1.00  &      0.059/0.117   &   I\\
HD 108147 b   &     38.57  &  1.05  &   0.340 &   0.0980 &   10.8810   &      0.5580 &  0.75/1.40    &    0.026/0.170   &   I\\
HD 38529 b &       42.43 &   1.39  &   0.770 &   0.1293  &  14.3200   &      0.2700 &  1.40/3.00    &    0.067/0.192  &    I(IV)\\
55 Cnc b       &   12.53  &   0.95  &   0.840 &   0.1150 &   14.6530 &        0.0200&   0.70/1.30 &       0.084/0.146   &   I (II-III)\\
Gl 86 b     &      11.00   &  0.79  &   4.000  &  0.1100  &  15.7800  &       0.0460 &  0.50/1.00  &      0.056/0.164  &    I\\
HD 195019 b   &    20.00  &   1.02   &  3.430  &  0.1400  &  18.3000   &      0.0500  & 0.70/1.30    &    0.079/0.201  &    I\\
HD 6434 b    &      40.32  &  1.00  &   0.480 &   0.1500 &   22.0900   &      0.3000  & 0.70/1.30   &     0.075/0.225  &    I\\
Gliese 876 c   &   04.69   &  0.32  &   0.560  &  0.1300  &  30.1200   &      0.2700   &0.10/0.20    &    0.054/0.206   &   III-IV(III-IV)\\
rho CrB b   &      16.70  &   0.95  &   1.100 &   0.2300 &   39.6450  &       0.0280 &  0.70/1.30   &     0.161/0.299  &    I\\
HD 74156 b  &      64.56 &   1.05 &    1.560  &  0.2760  &  51.6100  &       0.6490 &  0.75/1.40 &       0.016/0.536  &    II(IV)\\
HD 168443 b    &   33.00   &  1.01   &  7.200  &  0.2900  &  57.9000     &    0.5500  & 0.70/1.30  &      0.000/0.594  &    II(III-IV)\\
Gliese 876 b   &   04.69  &   0.32  &   1.890 &   0.2100  &  61.0200  &       0.1000  & 0.10/0.20    &    0.091/0.329   &   III-IV(III-IV)\\
HD 121504 b  &     44.37  &  1.00&     0.890&    0.3200&    64.6000  &       0.1300 &  0.70/1.30   &     0.199/0.441   &   II\\
HD 178911 b    &   46.73   & 0.87 &    6.292  &  0.3200  &  71.4870  &       0.1243  & 0.60/1.20    &    0.121/0.519  &    II-III\\
HD 16141 b     &    35.90 &    1.00 &    0.215 &   0.3500 &   75.8200    &     0.2800  & 0.70/1.30   &     0.198/0.502  &    II\\
HD 114762 b   &    40.57  &  0.82   &  10.990  & 0.3500  &  83.9090   &      0.3400  & 0.50/1.00  &      0.016/0.684  &    III\\
HD 80606 b     &   58.38  &  0.90   &  3.410  &  0.4390 &  111.7800  &       0.9270  & 0.60/1.20  &      0.000/1.023   &   III\\
70 Vir b   &       22.00  &   1.10  &   6.600  &  0.4300 &  116.0000   &      0.4000   &0.85/1.60    &    0.056/0.804  &    II-III\\
HD 52265 b  &      28.00  &   1.13  &   1.130 &   0.4900 &  118.9600   &      0.2900 &  0.85/1.60 &       0.221/0.759  &    II\\
GJ 3021 b  &       17.62  &  0.90  &   3.320  &  0.4900  & 133.8200   &      0.5050  & 0.60/1.20  &      0.047/0.933  &    III\\
HD 37124 b   &     33.20  &   0.91 &    0.860 &   0.5400 &  153.0000   &      0.1000 &  0.60/1.20     &   0.349/0.731   &   III(III-IV)\\
HD 82943 c    &    27.46  &  1.05  &   0.880 &   0.7300 &  221.6000    &     0.5400 &  0.75/1.40  &      0.168/1.302  &    III-IV(IV)\\
HD 8574 b      &   44.15  &  1.10   &  2.230  &  0.7600 &  228.8000   &      0.4000  & 0.85/1.60    &    0.208/1.312  &    III\\
HD 169830 b   &    36.32  &  1.40  &   2.940  &  0.8230  & 229.9000  &       0.3500  & 1.40/3.00  &      0.263/1.383   &   II-III\\
\\
\hline
\end{tabular}
\label{tab:one}
\end{center}
NOTE. -- (1) In order of increasing orbital period (3) Stellar mass
(4) Giant planet minimum mass (5) Semi-major axis (6) Orbital period
(7) Eccentricity (8) From Kasting et al. (1993) (9) See definition in
text (10) See definition in text.}
\end{table*}

\clearpage

\begin{table*}
\tablenum{1}
{\tiny
\caption{{\it Continued}}
\begin{center}
\begin{tabular}{lccccrcccc} \hline \hline
\\
Planet & Distance& $M_*$ &$M_p \sin i$ & $a$  & $P_{\rm orb}$ & $e$ & Habitable Zone &  Zone of Influence & Class\\
& (pc) & ($M_\odot$) & ($M_{\rm Jup}$) & (AU) & (days) & & (AU) & (AU)\\
(1)&(2)&(3)&(4)&(5)&(6)&(7)&(8)&(9)&(10)\\
\\
\hline
\\
Ups And c    &     16.50  &   1.30 &    2.110 &   0.8300 &  241.2000   &      0.1800 &  1.25/2.70    &    0.429/1.231   &   II-III(IV)\\
HD 89744 b   &     40.00  &   1.40   &  7.200  &  0.8800  & 256.0000     &    0.7000  & 1.40/3.00   &     0.000/1.888    &  III\\
HD 134987 b  &     25.00  &   1.05 &    1.580  &  0.7800 &  260.0000   &      0.2400 &  0.75/1.40 &       0.362/1.198  &    III\\
HD 12661 b     &   37.16  &   1.07   &  2.260  &  0.8200  & 260.8910    &     0.3500  & 0.80/1.45  &      0.261/1.379  &    III-IV (III-IV)\\
HR 810 (HD 17051) b  &      15.50  &   1.03   &  2.250  &  0.9250 &  320.1000    &     0.1610 &  0.70/1.30  &      0.466/1.384  &    IV*\\
HD 142 b     &     20.60 &    1.10 &    1.000  &  0.9800&   337.1120  &       0.3800 &  0.85/1.60 &       0.363/1.597  &    III-IV\\
HD 92788 b     &   32.82  &  1.06  &   3.830  &  0.9600  & 354.0000   &      0.3000  & 0.75/1.40  &      0.292/1.628   &   IV\\
HD 28185 b   &     39.40   &  0.99   &  5.700  &  1.0300  & 383.0000    &     0.0700  & 0.70/1.30   &     0.481/1.579  &    IV*\\
HD 177830 b   &    59.00 &    1.17 &    1.280 &   1.0000&   391.0000  &       0.4300 &  0.93/1.80 &       0.304/1.696 &     III-IV\\
HD 4203 b     &    77.50  &   1.06  &   1.640   & 1.0900  & 406.0000   &      0.5300  & 0.75/1.40  &      0.187/1.993  &    IV\\
HD 27442 b  &      18.20 &    1.20  &   1.280 &   1.1600 &  416.7040 &        0.0700 &  0.93/1.80  &      0.773/1.547  &    III*\\
HD 210277 b  &     22.00 &    0.92 &    1.240 &   1.0970  & 435.6000     &    0.4500  & 0.65/1.25 &       0.291/1.903  &    IV\\
HD 82943 b    &    27.46 &   1.05 &    1.630 &   1.1600 &  444.6000   &      0.4100 &  0.75/1.40 &       0.338/1.982   &   IV(IV)\\
HD 19994 b     &   22.38  &  1.35  &   2.000 &   1.3000  & 454.0000  &       0.2000  & 1.33/2.85  &      0.658/1.942  &    III\\
HD 114783 b  &     22.00 &    0.92   &  1.000  &  1.2000 &  501.0000    &     0.1000  & 0.65/1.25  &      0.762/1.638   &   III-IV\\
HIP 75458 b  &     31.50 &   1.05  &  8.640 &  1.3400 & 550.6510 &       0.7100 & 0.75/1.40 &      0.000/2.990   &  IV\\
HD 222582 b    &   42.00  &   1.00   &  5.400  &  1.3500 &  576.0000   &      0.7100  & 0.70/1.30  &      0.000/2.920  &    IV\\
HD 23079 b     &   34.80   &  1.10  &   2.540  &  1.4800 &  627.3000  &       0.0600  & 0.85/1.60  &      0.886/2.074  &    III-IV*\\
HD 141937 b    &   33.46  &  1.00   &  9.700  &  1.4900  & 658.8000   &      0.4040  & 0.70/1.30    &    0.068/2.912  &    IV\\
HD 160691 b    &   15.20  &   1.08   &  1.970  &  1.6500   & 743.0000   &      0.6200  & 0.85/1.60    &    0.107/3.193   &   IV\\
16 CygB b     &    21.40  &   1.01  &   1.500 &   1.7200 &  804.0000  &       0.6700 &  0.70/1.30  &      0.061/3.379  &    IV\\
HD 4208 b  &       33.90  &   0.93 &    0.810  &  1.6900 &  829.0000    &     0.0400 &  0.65/1.25 &       1.206/2.174  &    II-III\\
HD 213240 b    &   40.75  &  1.22  &   4.500  &  2.0300  & 951.0000     &    0.4500  & 0.93/1.80     &   0.307/3.753  &    IV\\
47 UMa b    &      13.30   &  1.03  &   2.540  &  2.0900 &  1089.0000   &     0.0610  & 0.70/1.30 &       1.234/2.946  &    II-III(II-III)\\
HD 10697 b     &   30.00   &  1.10  &   6.590   & 2.0000  & 1093.0000  &      0.1200  & 0.85/1.60 &       0.823/3.177   &   III-IV\\
HD 190228 b    &   66.11  &  1.30   &  4.990  &  2.3100  & 1127.0000  &      0.4300  & 1.25/2.70    &    0.383/4.237  &    IV\\
HD 136118 b	& 52.30	  &  1.24   & 12.050 &  2.4600   & 1259.9840       & 0.4000  & 1.10/2.25    &      0.121/4.799 & IV \\
Ups And d     &    16.50  &   1.30  &   4.610 &   2.5000 &  1266.6000  &      0.4100  & 1.25/2.70  &      0.491/4.509   &   IV(IV)\\
HD 50554 b    &    31.03  &  1.10   &  4.900 &   2.3800  & 1279.0000     &   0.4200  & 0.85/1.60  &      0.370/4.390    &  IV\\
HD 106252 b     &  37.44  &  1.05  &   6.810  &  2.6100  & 1500.0000   &     0.5400  & 0.75/1.40 &     0.000/5.275  &    IV\\
HD 33636 b     &   38.00  &   0.99  &   7.710  &  2.6200 &  1553.0000    &    0.3900  & 0.70/1.30 &       0.258/4.982  &    IV\\
14 Her (HD 145675) b  &     17.00   &  0.79  &   3.300  &  2.5000  & 1650.0000  &      0.3260  & 0.50/1.00   &     0.646/4.354   &   III-IV\\
HD 39091 b    &    20.55  &  1.10  &   10.370  & 3.3400  & 2115.3000   &     0.6200  & 0.85/1.60   &     0.000/7.232  &   IV\\
HD 168443 c    &   33.00   &  1.01   &  17.100  & 2.8700  & 2136.7000  &      0.2000  & 0.70/1.30   &    0.394/5.346   &   III-IV(III-IV)\\
HD 74156 c  &      64.56 &   1.05 &    7.500 &   3.4700 &  2300.0000    &    0.3950 &  0.75/1.40 &       0.375/6.565  &    IV(IV)\\
Epsilon Eridani b & 03.20   &   0.80  &   0.860 &   3.3000 &  2502.1000  &      0.6080 &  0.50/1.00   &     0.421/6.179 &     III-IV\\
47 UMa c      &    13.30  &   1.03  &   0.760  &  3.7300  & 2594.0000    &    0.1000  & 0.70/1.30  &      2.487/4.973  &    II(II-III)\\
\\
\hline
\end{tabular}
\end{center}
NOTE. -- (1) In order of increasing orbital period (3) Stellar mass
(4) Giant planet minimum mass (5) Semi-major axis (6) Orbital period
(7) Eccentricity (8) From Kasting et al. (1993) (9) See definition in
text (10) See definition in text.}
\end{table*}

\clearpage

\begin{table*}
{\tiny
\caption{23 RECENTLY DETECTED EXTRASOLAR GIANT PLANETS}
\begin{center}
\begin{tabular}{lccccrcccc} \hline \hline
\\
Planet & Distance& $M_*$ &$M_p \sin i$ & $a$  & $P_{\rm orb}$ & $e$ & Habitable Zone &  Zone of Influence & Class\\
& (pc) & ($M_\odot$) & ($M_{\rm Jup}$) & (AU) & (days) & & (AU) & (AU)\\
(1)&(2)&(3)&(4)&(5)&(6)&(7)&(8)&(9)&(10)\\
\\
\hline
\\
HD 49674 b & 40.70 & 1.00 & 0.12 & 0.060 & 4.94 & 0.0000 & 0.70/1.30 & 0.050/0.064 & I\\
55 Cnc c & 12.53 & 0.95 & 0.21 & 0.240 & 44.28 & 0.3400 & 0.70/1.30 & 0.121/0.359 & I(II-III)\\
HD 223084 b & 38.60 & 1.05 & 1.18 & 0.410 & 101.06 & 0.4800 & 0.75/1.40 & 0.103/0.717 & II-III\\
HD 73526 b & 94.60 & 1.02 & 3.03 & 0.647 & 186.90 & 0.4100 & 0.70/1.30 & 0.142/1.152 & III-IV\\
HD 150706 b & 27.20 & 1.00$^a$ & 1.00 & 0.820 & 264.90 & 0.3800 & 0.70/1.30 & 0.297/1.343 & IV\\
HD 40979 b & 33.00 & 1.08 & 3.32 & 0.811 & 267.20 & 0.2300 & 0.85/1.60 & 0.320/1.302 & III \\
HD 108874 b & 68.50 & 1.00 & 1.65 & 1.060 & 401.00 & 0.2000 & 0.70/1.30 & 0.525/1.595 & IV*\\
HD 128311 b & 16.60 & 0.80 & 2.63 & 1.010 & 414.00 & 0.2100 & 0.50/1.00& 0.410/1.610 & IV\\
HD 20367 b & 27.10 & 1.05 & 1.07 & 1.250 & 500.00 & 0.2300 & 0.75/1.40 & 0.638/1.862 & IV\\
HD 147513 b & 12.90 & 0.92 & 1.00 & 1.260 & 540.40 & 0.5200 & 0.65/1.25 & 0.270/2.250 & IV\\
HD 114386 b & 28.00 & 0.75 & 0.99 & 1.620 & 872.00 & 0.2800 & 0.50/1.00 & 0.708/2.532 & III-IV\\
HD 114729 b & 35.00 & 0.93 & 0.90 & 2.080 & 1136.00 & 0.3300 & 0.65/1.25 & 0.863/3.297 & III-IV\\
HD 196050 b & 46.90& 1.10 & 3.00 & 2.500 & 1288.00 & 0.2800 & 0.85/1.60 & 0.899/4.101 & III-IV\\
HD 216437 b & 26.50 & 1.07 & 2.09 & 2.700 & 1294.00 & 0.3400 & 0.80/1.45 & 0.911/4.489 & III-IV\\
HD 13507 b & 26.20 & 1.05 & 3.46 & 2.390 & 1318.00 & 0.1300 & 0.75/1.40 & 1.161/3.619 & III-IV\\
HD 12661 c & 37.16 & 1.07 & 1.66 & 2.610 & 1407.00 & 0.2240 & 0.80/1.45 & 1.299/3.821 & III-IV(III-IV)\\
HD 23596 b & 52.00 & 1.10 & 7.19 & 2.720 & 1558.00 & 0.3140 & 0.85/1.60 & 0.553/4.887 & IV\\
HD 30177 b & 54.70 & 0.95 & 7.95 & 2.650 & 1620.00 & 0.2168 & 0.70/1.30 & 0.687/4.613 & III-IV\\
HD 37124 c & 33.20 & 0.91 & 1.01 & 2.950 & 1942.00 & 0.4000 & 0.60/1.20 & 0.982/4.918 & III-IV(III-IV)\\
HD 72659 b & 51.40 & 0.95 & 2.55 & 3.240 & 2185.00 & 0.1800 & 0.70/1.30 & 1.495/4.985 & II-IV\\
HD 38529 c & 42.43 & 1.39 & 11.30 & 3.510 & 2189.53 & 0.3400 & 1.40/3.00 & 0.495/6.525 & IV(IV)\\
Gl 777A (HD 190360A) b & 15.90 & 0.90 & 1.15 & 3.650 & 2613.00 & 0.0000 & 0.60/1.20 & 2.628/4.672 & II\\
55 Cnc d  & 12.53 & 0.95 & 4.00 & 5.900 & 5360.00 & 0.1600 & 0.70/1.30 & 2.487/9.313 & II-III (II-III)\\
\\
\hline
\end{tabular}
\label{tab:onebis}
\end{center}
NOTE. -- (1) In order of increasing orbital period (3) Stellar mass
(4) Giant planet mass (5) Semi-major axis (6) Orbital period (7)
Eccentricity (8) From Kasting et al. (1993) (9) See definition in text
(10) See definition in text (a) Assumed given $L_* \simeq L_\odot$.}
\end{table*}

\clearpage
\begin{table*}
\caption{DETAILED ORBITAL PARAMETERS FOR 3 EXTRASOLAR GIANT PLANETS}
\begin{center}
\begin{tabular}{lccrc} \hline \hline
\\
Planet &  $M_*$ &$M_p \sin i$ &  $P_{\rm orb}$ & $e$ \\
& ($M_\odot$) & ($M_{\rm Jup}$) & (days) & \\
(1)&(2)&(3)&(4)&(5)\\
\\
\hline
\\
HD 169830 b$^a$   &   1.40 ($\pm$ 0.21)  &   2.94 $\pm$ 0.12  & 229.9
$\pm$ 4.6  &       0.35 $\pm$ 0.04 \\
HD 114783 b$^b$  &     0.92 ($\pm$ 0.138)  &  1.0   &   501.0
$\pm$ 14    &     0.10 $\pm$ 0.08 \\
HD 210277 b$^c$  &    0.92 ($\pm$ 0.138)&    1.24 $\pm$ 0.03 &  435.6
$\pm$ 1.9     &    0.450 $\pm$ 0.015 \\
\\
\hline
\end{tabular}
\label{tab:two}
\end{center}
NOTE. -- (2) Stellar mass; error is an arbitrary $15\%$ (3) Giant
planet minimum mass (4) Orbital period (5) Eccentricity (a) Naef et
al. 2001 (b) Vogt et al. 2002 (c) Naef et al. 2001 -- ``combined
solution''.
\end{table*}

\clearpage
\begin{table*}
\caption{RESULTS FOR 3 TEST SYSTEMS}
\begin{center}
\begin{tabular}{lccc} \hline \hline
\\
Planet &  Class & Remaining
& Remaining \\
& & (Exp. I) & (Exp. II)\\
\\
\hline
\\
HD 169830 b & II & 29/100 & 27/100 \\
HD 114783 b & III & 6/100 & 1/100 \\
HD 210277 b & IV & 0/100 & 0/100 \\
\\
\hline
\end{tabular}
\label{tab:twobis}
\end{center}
\end{table*}

\clearpage

\begin{table*}
{\tiny
\caption{STRONG MEAN--MOTION RESONANCES IN HABITABLE ZONES (CLASS I \& II)}
\begin{center}
\begin{tabular}{lcrcc} \hline \hline
\\
Planet & $a$  & $P_{\rm orb}$ & Habitable Zone &  Resonances\\
&  (AU) & (days) & (AU) & (AU)\\
(1)&(2)&(3)&(4)&(5)\\
\\
\hline
\\
HD 83443 b  &        0.0380  &   2.9853  &       0.50/1.00    &  --   \\
HD 46375 b   &         0.0410  &   3.0240     &    0.70/1.30  &  --    \\
HD 187123 b     &      0.0420 &    3.0900    &    0.75/1.40  &   --    \\
HD 179949 b   &       0.0450  &   3.0930     &    1.10/2.25   &  --    \\
Tau Boo b     &       0.0500   &  3.3120  &       1.25/2.70  &   --    \\
BD -10316 b   &       0.0460  &   3.4870   &      0.85/1.60  &   --    \\
HD 75289 b     &      0.0460  &   3.5100    &     0.75/1.40  &   --    \\
HD 209458 b    &       0.0450  &   3.5247  &      0.75/1.40  &   --    \\
51 Peg b    &          0.0512  &   4.2300   &     0.70/1.30    &  --   \\
HD 68988 b    &       0.0710   &  6.2760   &      0.93/1.80     &  --  \\
HD 168746 b   &       0.0660 &    6.4090   &      0.65/1.25     &  --  \\
HD 217107 b   &       0.0700  &   7.1270     &    0.70/1.30 &   --     \\
HD 130322 b   &       0.0880  &  10.7240    &     0.50/1.00  &  --     \\
HD 108147 b   &       0.0980 &   10.8810   &      0.75/1.40    &  --   \\
HD 38529 b &         0.1200  &  14.3007   &      1.40/3.00    &  --     \\
55 Cnc b       &      0.1150 &   14.6530 &        0.70/1.30 &     --    \\
HD 195019 b   &       0.1400  &  18.3000   &      0.70/1.30    &  --   \\
HD 6434 b    &        0.1500 &   22.0900   &      0.70/1.30   &  --    \\
rho CrB b   &         0.2300 &   39.6450  &       0.70/1.30   &  --    \\
HD 121504 b  &       0.3200&    64.6000  &       0.70/1.30   &   0.806 (4:1)   \\
HD 178911 b    &     0.3200  &  71.4870  &       0.60/1.20    &  0.666 (3:1) / 0.806 (4:1)   \\
HD 16141 b     &       0.3500 &   75.8200    &    0.70/1.30   &   0.728 (3:1) / 0.882 (4:1)  \\
70 Vir b   &          0.4300 &  116.0000   &     0.85/1.60    &   0.894 (3:1) / 1.084 (4:1)  \\
HD 52265 b  &         0.4900 &  118.9600   &      0.85/1.60 &     0.903 (5:2) / 1.019 (3:1) / 1.235 (4:1)  \\
HD 169830 b   &      0.8230  & 229.9000  &       1.40/3.00  &     1.516 (5:2) / 1.712 (3:1) / 2.074 (4:1)  \\
HD 4208 b  &          1.6900 &  829.0000    &     0.65/1.25 &     0.812 (1:3) / 1.065 (1:2) / 1.202 (3:5) / ...   \\
47 UMa b    &         2.0900 &  1089.0000   &     0.70/1.30 &      0.829 (1:4) / 1.005 (1:3) / 1.135 (2:5)  \\
47 UMa c      &       3.7300  & 2594.0000    &    0.70/1.30  &     --  \\
Gl 777A (HD 190360A) b & 3.6500 & 2613.0000 & 0.60/1.20 & --\\
\\
\hline
\end{tabular}
\label{tab:threebis}
\end{center}
NOTE. -- (1) In order of increasing orbital period (2) Semi-major axis
(3) Orbital period (4) From Kasting et al. (1993) (5) Mean-motion
resonances of highest order ($p/p+q$, $q=1,2,3$)}
\end{table*}

\clearpage

\begin{table*}
{\tiny
\caption{REMAINING HABITABLE TEST PARTICLES AFTER $10^6$ YEARS}
\begin{center}
\begin{tabular}{lccccccccc} \hline \hline
\\
System/Planets & $\sin i$ = 0.9 & $\sin i$ = 0.7 & $\sin i$ = 0.5 & $\sin i$ =
0.3 & $\sin i$ = 0.1 & Average & $a$ (AU) & $e$ & $i$ (deg) \\ 
\\
\hline
\\
Equiv. Solar System & 73 & 71 & 76 & 81 & 73 & 74.8 & $0.98\pm0.13$ & $0.06\pm0.03$ & $6.3\pm3.5$ \\  
HD 83443 b	& 83 & 78 & 75 & 76 & 70 & 76.4 & $0.74\pm0.11$ & $0.06\pm0.03$ & $6.3\pm3.3$ \\
HD 46375 b	& 79 & 79 & 72 & 79 & 85 & 78.8 & $0.99\pm0.13$ & $0.06\pm0.03$ & $6.5\pm3.2$ \\
HD 187123 b	& 80 & 79 & 72 & 80 & 78 & 77.8 & $1.08\pm0.15$ & $0.06\pm0.03$ & $6.5\pm3.4$ \\
HD 179949 b	& 78 & 82 & 88 & 80 & 71 & 79.8 & $1.66\pm0.28$ & $0.07\pm0.04$ & $6.5\pm3.3$ \\
Tau Boo b   	& 77 & 80 & 73 & 70 & 29 & 65.8 & $1.96\pm0.40$ & $0.11\pm0.09$ & $6.7\pm3.6$ \\
BD -10316 b	& 82 & 89 & 81 & 82 & 75 & 81.8 & $1.22\pm0.18$ & $0.06\pm0.03$ & $6.5\pm3.3$ \\
HD 75289 b	& 84 & 84 & 76 & 78 & 78 & 80.0 & $1.07\pm0.15$ & $0.06\pm0.03$ & $6.6\pm3.4$ \\
HD 209458 b	& 82 & 73 & 78 & 76 & 72 & 76.2 & $1.04\pm0.15$ & $0.06\pm0.03$ & $6.4\pm3.3$ \\
51 Peg b	& 75 & 75 & 78 & 84 & 81 & 78.6 & $0.99\pm0.14$ & $0.06\pm0.03$ & $6.8\pm3.7$ \\
Ups And b,c,d	&  0 &  0 &  0 &  0 &  0 &   0.0 & -- & -- & -- \\
HD 49674 b      & 78 & 80 & 81 & 75 & 76 & 78.0 & $0.99\pm0.15$ & $0.06\pm0.03$ & $6.3\pm3.5$ \\
HD 68988 b	& 84 & 69 & 71 & 69 & 58 & 70.2 & $1.34\pm0.19$ & $0.08\pm0.05$ & $6.7\pm3.4$ \\
HD 168746 b	& 80 & 74 & 81 & 84 & 81 & 80.0 & $0.93\pm0.14$ & $0.06\pm0.03$ & $6.7\pm3.5$ \\
HD 217107 b	& 74 & 75 & 71 & 74 & 64 & 71.6 & $0.99\pm0.13$ & $0.07\pm0.04$ & $7.0\pm3.8$ \\
HD 130322 b	& 74 & 84 & 80 & 79 & 71 & 77.6 & $0.74\pm0.12$ & $0.07\pm0.04$ & $6.7\pm3.3$ \\
HD 108147 b	& 59 & 72 & 78 & 71 & 0  & 56.0 & $1.07\pm0.14$ & $0.07\pm0.03$ & $6.9\pm3.6$ \\
HD 38529 b,c    & 0 & 0 & 0 & 0 & 0 & 0.0 & -- & -- & -- \\
55 Cnc b,c,d    & 67 & 0 & 0 & 0 & 0 & 13.4 & $1.00\pm0.10$ & $0.09\pm0.04$ & $6.9\pm3.7$ \\
Gl 86 b		& 77 & 84 & 70 & 69 & 34 & 66.8 & $0.75\pm0.12$ & $0.09\pm0.07$ & $6.3\pm3.4$ \\
HD 195019 b	& 65 & 69 & 73 & 65 & 45 & 63.4 & $1.01\pm0.13$ & $0.08\pm0.05$ & $6.6\pm3.4$ \\
HD 6434 b	& 53 & 51 & 39 & 58 & 52 & 50.6 & $0.99\pm0.11$ & $0.09\pm0.04$ & $6.7\pm3.7$ \\
Gliese 876 b,c	&  0 &  0 &  0 &  0 &  0 &   0.0 & -- & -- & --  \\
rho CrB b	& 80 & 75 & 72 & 76 & 76 & 75.8 & $1.00\pm0.14$ & $0.06\pm0.03$ & $6.6\pm3.6$  \\
HD 74156 b,c	&  0 &  0 &  0 &  0 &  0 &   0.0 & -- & -- & --  \\
HD 168443 b,c	&  0 &  0 &  0 &  0 &  0 &   0.0 & -- & -- & --  \\
HD 121504 b	& 58 & 55 & 53 & 58 & 60 & 56.8 & $0.99\pm0.11$ & $0.08\pm0.04$ & $6.0\pm3.3$ \\
HD 178911 b	& 63 & 46 & 50 & 42 & 11 & 42.4 & $0.90\pm0.12$ & $0.11\pm0.05$ & $6.7\pm3.3$ \\
HD 16141 b	& 26 & 22 & 19 & 23 & 28 & 23.6 & $1.00\pm0.06$ & $0.14\pm0.06$ & $7.2\pm3.5$ \\
HD 114762 b	&  1 &  2 &  0 &  0 &  0 &  0.6 & $0.75\pm0.05$ & $0.13\pm0.06$ & $6.3\pm2.8$ \\ 
HD 223084 b     & 9 & 8 & 5 & 3 & 0 & 5.0 & $1.06\pm0.07$ & $0.13\pm0.06$ & $6.4\pm3.7$ \\
HD 80606 b	&  0 &  0 &  0 &  0 &  0 &   0.0 & -- & -- & --  \\
70 Vir b	&  7 &  5 &  3 &  0 &  0 &  3.0 & $1.23\pm0.06$ & $0.17\pm0.08$ & $5.7\pm3.0$ \\
HD 52265 b	& 17 & 13 & 22 & 14 &  7 & 14.6 & $1.19\pm0.08$ & $0.13\pm0.06$ & $6.5\pm3.2$ \\
GJ 3021 b	&  0 &  0 &  0 &  0 &  0 &   0.0 & -- & -- & --  \\
HD 37124 b,c    & 0 & 0 & 0 & 0 & 0 & 0.0 & -- & -- & -- \\
HD 73526 b      & 1 & 0 & 0 & 0 & 0 & 0.2 & $1.03\pm0.0$ & $0.09\pm0.0$ & $5.2\pm0.0$   \\
HD 82943 b,c	&  0 &  0 &  0 &  0 &  0 &   0.0  & -- & -- & -- \\
HD 8574 b	&  0 &  0 &  0 &  0 &  0 &   0.0 & -- & -- & --  \\
HD 169830 b	& 22 & 23 & 26 & 22 &  9 & 20.4 & $2.15\pm0.16$ & $0.16\pm0.07$ & $6.8\pm4.2$ \\
HD 89744 b	&  0 &  0 &  0 &  0 &  0 &   0.0  & -- & -- & -- \\
HD 134987 b	&  0 &  0 &  0 &  0 &  0 &   0.0 & -- & -- & --  \\
HD 12661 b,c	&  0 &  0 &  0 &  0 &  0 &   0.0 & -- & -- & --  \\
HD 150706 b     &  0 &  0 &  0 &  0 &  0 & 0.0 & -- & -- & -- \\
HD 40979 b      &  0 &  0 &  0 &  0 &  0 & 0.0 & -- & -- & -- \\
HR 810 (HD 17051) b  &  0 &  0 &  0 &  0 &  0 &   0.0 & -- & -- & --  \\
\\
\hline
\end{tabular}
\label{tab:three}
\end{center}}
\end{table*}

\clearpage

\begin{table*}
\tablenum{6}
{\tiny
\caption{{\it Continued}}
\begin{center}
\begin{tabular}{lccccccccc} \hline \hline
\\
System/Planets & $\sin i$ = 0.9 & $\sin i$ = 0.7 & $\sin i$ = 0.5 & $\sin i$ =
0.3 & $\sin i$ = 0.1 & Average & $a$ (AU) & $e$ & $i$ (deg) \\
\\
\hline
\\
HD 142 b	&  0 &  0 &  0 &  0 &  0 &   0.0  & -- & -- & -- \\
HD 92788 b	&  0 &  0 &  0 &  0 &  0 &   0.0 & -- & -- & --  \\
HD 28185 b	&  3 &  2 &  0 &  0 &  0 &   1.0 & $1.03\pm0.03$ & $0.11\pm0.03$ & $9.6\pm3.2$  \\
HD 177830 b	&  1 &  0 &  0 &  0 &  0 &  0.2 & $1.59\pm0.0$ & $0.02\pm0.0$ & $9.7\pm0.0$  \\
HD 108874 b     &  0 &  0 &  0 &  0 &  0 & 0.0 & -- & -- & -- \\
HD 4203 b	&  0 &  0 &  0 &  0 &  0 &   0.0 & -- & -- & --  \\
HD 128311 b     &  0 &  0 &  0 &  0 &  0 & 0.0 & -- & -- & -- \\
HD 27442 b	&  7 &  8 &  5 &  3 &  2 &   5.0 & $1.31\pm0.18$ & $0.08\pm0.04$ & $5.8\pm2.8$  \\
HD 210277 b 	&  0 &  0 &  0 &  0 &  0 &   0.0 & -- & -- & --  \\
HD 19994 b	&  1 &  1 &  1 &  0 &  1 &  0.8 & $2.26\pm0.18$ & $0.18\pm0.18$ & $6.2\pm5.8$ \\
HD 20367 b      &  0 &  0 &  0 &  0 &  0 & 0.0 & -- & -- & -- \\
HD 114783 b	&  1 &  0 &  0 &  0 &  0 &  0.2 & $0.98\pm0.0$ & $0.25\pm0.0$ & $8.8\pm0.0$  \\
HD 147513 b     &  0 &  0 &  0 &  0 &  0 & 0.0 & -- & -- & -- \\
HIP 75458 b	&  0 &  0 &  0 &  0 &  0 &   0.0  & -- & -- & -- \\
HD 222582 b	&  0 &  0 &  0 &  0 &  0 &   0.0  & -- & -- & -- \\
HD 23079 b	&  0 &  0 &  0 &  0 &  0 &   0.0  & -- & -- & -- \\
HD 141937 b	&  0 &  0 &  0 &  0 &  0 &   0.0  & -- & -- & -- \\
HD 160691 b	&  0 &  0 &  0 &  0 &  0 &   0.0  & -- & -- & -- \\
16 CygB b	&  0 &  0 &  0 &  0 &  0 &   0.0  & -- & -- & -- \\
HD 4208 b	& 67 & 65 & 59 & 47 & 13 & 50.2 & $0.90\pm0.12$ & $0.07\pm0.04$ & $6.3\pm3.2$  \\
HD 114386 b     & 11 & 7 & 4 & 0 & 0 & 4.4 & $0.71\pm0.05$ & $0.17\pm0.06$ & $7.8\pm4.9$  \\
HD 213240 b	&  0 &  0 &  0 &  0 &  0 &   0.0  & -- & -- & -- \\
47 UMa b,c	& 34 & 16 & 50 & 40 &  0 & 28.0 & $0.99\pm0.10$ & $0.08\pm0.04$ & $6.6\pm3.3$  \\
HD 10697 b	&  0 &  0 &  0 &  0 &  0 &   0.0  & -- & -- & -- \\
HD 190228 b	&  0 &  0 &  0 &  0 &  0 &   0.0 & -- & -- & -- \\
HD 114729 b     & 0 & 0 & 0 & 0 & 0 & 0.0 & -- & -- & -- \\
HD 136118 b	&  0 &  0 &  0 &  0 &  0 &   0.0  & -- & -- & -- \\
HD 50554 b	&  0 &  0 &  0 &  0 &  0 &   0.0  & -- & -- & -- \\
HD 196050 b     & 1 & 0 & 0 & 0 & 0 & 0.2 & $0.98\pm0.0$ & $0.05\pm0.0$ & $6.2\pm0.0$ \\ 
HD 216437 b     & 2 & 0 & 0 & 0 & 0 & 0.4 & $1.05\pm0.0$ & $0.15\pm0.06$ & $6.6\pm1.1$  \\
HD 13507 b      & 49 & 48 & 38 & 10 & 0 & 29.0 & $0.99\pm0.09$ & $0.09\pm0.04$ & $6.0\pm3.2$  \\
HD 106252 b	&  0 &  0 &  0 &  0 &  0 &   0.0  & -- & -- & -- \\
HD 33636 b	&  0 &  0 &  0 &  0 &  0 &   0.0  & -- & -- & -- \\
HD 23596 b      & 0 & 0 & 0 & 0 & 0 & 0.0 & -- & -- & -- \\
HD 30177 b      & 10 & 0 & 0 & 0 & 0 & 2.0 & $0.89\pm0.04$ & $0.09\pm0.03$ & $6.4\pm2.4$  \\
14 Her (HD 145675) b  & 26 & 16 &  4 &  0 &  0 & 9.2 & $0.70\pm0.06$ & $0.14\pm0.06$ & $6.1\pm3.4$ \\
HD 39091 b	&  0 &  0 &  0 &  0 &  0 &   0.0  & -- & -- & -- \\
HD 72659 b      & 60 & 48 & 52 & 41 & 0 & 40.2 & $0.95\pm0.10$ & $0.10\pm0.05$ & $6.9\pm3.5$  \\
Epsilon Eridani b &  0 &  0 &  0 &  0 &  0 &   0.0  & -- & -- & -- \\
GL 777A (HD 190360A) b & 83 & 89 & 91 & 87 & 84 & 86.8 & $0.88\pm0.14$ & $0.06\pm0.03$ & $6.2\pm3.2$  \\
\\
\hline
\end{tabular}
\end{center}}
\end{table*}

\clearpage

\begin{table*}
{\tiny
\caption{GLOBAL STATISTICAL RESULTS BY DYNAMICAL CLASS}
\begin{center}
\begin{tabular}{lcccccc} \hline \hline
\\
Class & \% Remaining & \% Close encounters & \# of remaining & $a$ (AU) &
$e$ & $i$ (deg)\\ 
& & (Leaving) & planets & (Remaining) & (Remaining)& (Remaining)\\
\\
\hline
\\
I & $73.9 \pm 8.3$\% & 0\% & 6926 & $1.08\pm0.34$ & $0.07\pm0.04$ &
$6.6^\circ\pm3.5^\circ$ \\
II & $29.9 \pm 17.3$\% & 0\% & 1449 & $1.01\pm0.29$ & $0.09\pm0.06$ &
$6.4^\circ\pm3.3^\circ$ \\
III & $0.92 \pm 2.31$\% & 50\% & 478 & $1.03\pm0.36$ & $0.10\pm0.06$ &
$6.4^\circ\pm3.2^\circ$ \\
IV & $0$\% & 85\% & 0 & -- & -- & -- \\       
\\
\hline
\end{tabular}
\label{tab:four}
\end{center}}
\end{table*}

\clearpage

\clearpage

\clearpage

\clearpage

\clearpage

\clearpage

\clearpage

\begin{thebibliography}{}
\bibitem[]{1261} Barnes, J.W. \& O'Brien, D.P. 2002, ApJ, in press,
astroph/0205035
\bibitem[]{1263} Beckwith, S.V.W. \& Sargent, A.I. 1996, Nature, 383, 139
\bibitem[]{1264} Charbonneau, D., Brown, T.M., Latham, D.W. \& Mayor, M.
2000, ApJ, 529, L45
\bibitem[]{1266} Charbonneau, D., Brown, T.M., Noyes, R.W. \& Gilliland,
R.L.  2002, ApJ, 568, 377
\bibitem[]{1268} Chiang, E.I., Tabachnik, S. \& Tremaine, S. 2001, AJ,
122, 1607
\bibitem[]{} Cumming, A., Marcy, G.W. , Butler, R.P. \& Vogt,
S.S. 2002, astro-ph/0209199
\bibitem[]{1270} Duncan, M., Quinn, T. \& Tremaine, S. 1987, AJ, 94, 1330
\bibitem[]{1271} Forget, F. \& Pierrehumbert, R.T. 1997, Science, 278,
1273
\bibitem[]{1273} Gaudi, B.S. 2002, in ASP Conference Series: Scientific
Frontiers in Research on Extrasolar Planets, eds. D. Deming and
S. Seager, astroph/0207533.
\bibitem[]{1276} Gehman, C.S., Adams, F.C. \& Laughlin, G. 1996, PASP,
108, 1018
\bibitem[]{1278} Hansen, B.M.S. 2002, ApJ Lett., submitted (astroph/0004058)
\bibitem[]{1279} Hart, M.H. 1978, Icarus, 33, 23
\bibitem[]{} Hays, J.D., Imbrie, J. \& Shackleton, N.J. 1976, Science,
194, 1121
\bibitem[]{1280} Henry, G.W., Marcy, G.W., Butler, R.P. \& Vogt,
S.S. 2000, ApJ, 529, L41
\bibitem[]{1282} Jones, B.W., Sleep, P.N. \& Chambers, J.E. 2001, A\&A,
366, 254
\bibitem[]{1284} Joshi, M.M., Haberle, R.M. \& Reynolds, R.T. 1997,
Icarus, 129, 450
\bibitem[]{1286} Kasting, J.F., Toon, O.B. \& Pollack, J.B. 1988,
Sci. Am., 258, 90
\bibitem[]{1288} Kasting, J.F., Whitmire, D.P. \& Reynolds, R.T. 1993,
Icarus, 101, 108
\bibitem[]{1290} Kozai, Y. 1962, AJ, 67, 591
\bibitem[]{1291} Laughlin, G., Chambers, J. \& Fischer, D. 2002, ApJ, in
press, astro-ph/0205514
\bibitem[]{1293} Marcy, G.W., Cochran, W.D. \& Mayor, M. 2000, in
Protostars and Planets IV, eds. Mannings, V., Boss, A.P., Russell,
S.S. (Tucson: University of Arizona Press), p. 1285
\bibitem[]{1296} Marcy, G.W. et al. 2002, ApJ, submitted (preprint)
\bibitem[]{1297} Mayor, M. et al. 2002, announcement during the "Scientific Frontiers in Research on Extrasolar Planets" conference, Washington D.C., June 19th, 2002.
\bibitem[]{1298} Mazeh, T. et al. 2000, ApJ, 532, L55
\bibitem[]{1299} Murray, C.D. \& Dermott, S.F. 1999, Solar System Dynamics
(Cambridge: Cambridge University Press)
\bibitem[]{1301} Murray, N., Hansen, B., Holman, M. \& Tremaine, S. 1998,
Science, 279, 69
\bibitem[]{1303} Naef, D. et al. 2001, A\&A, 375, 205
\bibitem[]{1304} North, G.R., Mengel, J.G. \& Short, D.A. 1983,
J. Geophys. Res., 88, 6576
\bibitem[]{1306} Rampino, M.R. \& Caldeira, K. 1994, ARA\&A, 32, 83
\bibitem[]{1307} Rivera, E.J. \& Lissauer, J.J. 2000, ApJ, 530, 454
\bibitem[]{1308} Rivera, E.J. \& Lissauer, J.J. 2001, ApJ, 558, 392 
\bibitem[]{1309} Ruden, S.P. 1999, in ``The Origin of Stars and Planetary
Systems.,'' Eds. C.J. Lada \& N.D. Kylafis (Kluwer Academic
Publishers), p. 643 (astroph/9910331)
\bibitem[Saha \& Tremaine, 1992]{sah92} Saha P., Tremaine S. 1994, AJ,
104, 1633 
\bibitem[Saha \& Tremaine, 1994]{sah94} Saha P., Tremaine S. 1994, AJ,
108, 1962
\bibitem[]{1316} Tabachnik, S. \& Tremaine, S. 2002, ApJ, in press,
astroph/0107482
\bibitem[]{1318} Vogt, S.S. et al. 2002, ApJ, 568, 352
\bibitem[]{1319} Williams, D.M., Kasting, J.F. \& Wade, R.A. 1997, Nature,
385, 234
\bibitem[]{1321} Williams, D.M. \& Kasting, J.F. 1997, Icarus, 129, 254
\bibitem[Wisdom \& Holman 1991]{wis91} Wisdom J., Holman M.J. 1991,
AJ, 102, 1528

\end{thebibliography}
\end{document}